\documentclass[longauth]{aa}  

\usepackage{graphicx}
\usepackage{txfonts}
\usepackage{graphicx}
\usepackage{txfonts}
\usepackage{graphicx}
\usepackage{lscape}
\usepackage{longtable}
\usepackage{natbib}
\usepackage{color}
\usepackage{array} 
\usepackage{array} 
\usepackage{tikz,array}
\usetikzlibrary{calc}
\usepackage{txfonts}
\usepackage{color}
\usepackage{multirow}
\usepackage{here}

\usepackage{txfonts}
\usepackage{amsmath,amstext}
\usepackage{graphicx}
\usepackage{epstopdf}
\usepackage{float}
\usepackage{array}
\usepackage{mathtools}
\usepackage{booktabs}
\usepackage{subfigure}
\usepackage{url}
\usepackage{helvet}
\usepackage{tabularx}
\usepackage{multirow}
\usepackage{natbib}
\usepackage[flushleft]{threeparttable}
\usepackage{lscape}
\usepackage{pdflscape}
\usepackage{longtable}
\usepackage{wasysym}
\usepackage{float}
\usepackage{relsize}
\usepackage{color}
\usepackage{breqn}
\usepackage{bm}
\usepackage{enumitem}
\usepackage{makecell}

\usepackage[colorlinks, citecolor=blue, linkcolor=blue]{hyperref}
\hypersetup{colorlinks,breaklinks, linkcolor=blue,urlcolor=magenta, anchorcolor=blue,citecolor=blue}

\def\kms{km s$^{-1}$}         
\def\ms{\hbox{m s$^{-1}$}}         
\def\gcm3{\hbox{g cm$^{-3}$}}       
\def\vsini{\hbox{$\upsilon \sin i_{\star}$}}      
\def\Msun{\hbox{$\mathrm{M}_{\astrosun}$}}             
\def\Rsun{\hbox{$\mathrm{R}_{\astrosun}$}}
\def\Mjup{\hbox{$\mathrm{M}_{\rm Jup}$}}

\def\Mearth{\hbox{$\mathrm{M}_{\oplus}$}}
\def\Rearth{\hbox{$\mathrm{R}_{\oplus}$}}

\bibpunct{(}{)}{;}{a}{}{,} 

\newcommand{\be}{\begin{equation}}
\newcommand{\ee}{\end{equation}}

\newcommand{\toi}{TOI-969}
\newcommand{\toib}{TOI-969\,b}
\newcommand{\toic}{TOI-969\,c}

\usepackage{romannum}

\begin{document}

   \title{TOI-969: a late-K dwarf with a hot mini-Neptune in the desert \\ and an eccentric cold Jupiter
   \thanks{Based on Guaranteed Time Observations collected at the European Southern Observatory (ESO) under ESO programs 1102.C-0249 (PI: Armstrong), 106.21TJ.001 (PI: Gandolfi). This paper includes data gathered with the 6.5 meter Magellan Telescopes located at Las Campanas Observatory, Chile. The full dataset is accessible from CDS through the following link XXX.}}


   \author{
J.~Lillo-Box\inst{\ref{cab}}, 
D.~Gandolfi\inst{\ref{torino}},
D.J.~Armstrong\inst{\ref{warwick},\ref{ceh}},
K.~A.~Collins\inst{\ref{cfa}},
L.~D.~Nielsen\inst{\ref{eso-garching},\ref{geneva}},
R.~Luque\inst{\ref{iaa}},
J.~Korth\inst{\ref{chalmers}},
S.~G.~Sousa\inst{\ref{ia-porto}},
S.~N.~Quinn\inst{\ref{cfa}},
L.~Acu\~na\inst{\ref{marseille}},
S.~B.~Howell\inst{\ref{ames}},
G.~Morello\inst{\ref{iac},\ref{ull}},
C.~Hellier\inst{\ref{keele}},
S.~Giacalone\inst{\ref{berkeley}},
S.~Hoyer\inst{\ref{marseille}},
K.~Stassun\inst{\ref{nashville}},
E.~Palle\inst{\ref{iac},\ref{ull}},
A.~Aguichine\inst{\ref{marseille}},
O.~Mousis\inst{\ref{marseille}},
V.~Adibekyan\inst{\ref{ia-porto},\ref{uporto}},
T.~Azevedo~Silva\inst{\ref{ia-porto},\ref{uporto}},
D.~Barrado\inst{\ref{cab}},
M.~Deleuil\inst{\ref{marseille}},
J.~D.~Eastman\inst{\ref{cfa}},
F.~Hawthorn\inst{\ref{warwick},\ref{ceh}},
J.~M.~Irwin\inst{\ref{ioa}},
J.~M.~Jenkins\inst{\ref{ames}},
D.~W.~Latham\inst{\ref{cfa}},
A.~Muresan\inst{\ref{chalmers2}},
C.M.~Persson\inst{\ref{chalmers2}},
A.~Santerne\inst{\ref{marseille}},
N.~C.~Santos\inst{\ref{ia-porto},\ref{uporto}},
A.~B.~Savel\inst{\ref{maryland}},
H.~P.~Osborn\inst{\ref{csh}},
J.~Teske\inst{\ref{carnegie}},
P.~J.~Wheatley\inst{\ref{warwick},\ref{ceh}},
J.~N.~Winn\inst{\ref{princeton}},
S.~C.~C.~Barros\inst{\ref{ia-porto},\ref{uporto}},
R.~P.~Butler\inst{\ref{carnegie}},
D.~A.~Caldwell\inst{\ref{seti}},
D.~Charbonneau\inst{\ref{cfa}},
R.~Cloutier\inst{\ref{cfa}},
J.~D.~Crane\inst{\ref{carnegie-ca}},
O.~D.~S.~Demangeon\inst{\ref{ia-porto},\ref{uporto}},
R.~F.~D\'iaz\inst{\ref{icas}},
X.~Dumusque\inst{\ref{geneva-inst}},
M.~Esposito\inst{\ref{tautenburg}},
B.~Falk\inst{\ref{stsci}},
H.~Gill\inst{\ref{berkeley}},
S.~Hojjatpanah\inst{\ref{marseille}},
L.~Kreidberg\inst{\ref{mpia}},
I.~Mireles\inst{\ref{newmex}},
A.~Osborn\inst{\ref{warwick},\ref{ceh}},
G.~R.Ricker\inst{\ref{mit}},
J.~E.~Rodriguez\inst{\ref{michigan}},
R.~P.~Schwarz\inst{\ref{cfa}},
S.~Seager\inst{\ref{mit},\ref{mit-earth},\ref{mit-aero}},
J.~Serrano Bell\inst{\ref{laplata}},
S.~A.~Shectman\inst{\ref{carnegie-ca}},
A.~Shporer\inst{\ref{mit}},
M.~Vezie\inst{\ref{mit}},
S.~X.~Wang\inst{\ref{tsinghua}},
G.~Zhou\inst{\ref{queensland}}
}

\institute{
Centro de Astrobiolog\'ia (CAB, CSIC-INTA), Depto. de Astrof\'isica, ESAC campus, 28692, Villanueva de la Ca\~nada (Madrid), Spain \email{Jorge.Lillo@cab.inta-csic.es}
\label{cab}
\and
Dipartimento di Fisica, Universit\'a degli Studi di Torino, via Pietro Giuria 1, 10125, Torino, Italy
\label{torino}
\and
Department of Physics, University of Warwick, Gibbet Hill Road, Coventry CV4 7AL, UK
\label{warwick}
\and
Centre for Exoplanets and Habitability, University of Warwick, Gibbet Hill Road, Coventry CV4 7AL, UK
\label{ceh}
\and
Center for Astrophysics \textbar \ Harvard \& Smithsonian, 60 Garden Street, Cambridge, MA 02138, USA
\label{cfa}
\and
European Southern Observatory, Karl-Schwarzschildstr. 2, D-85748 Garching bei M{\"u}nchen, Germany
\label{eso-garching}
\and
Observatoire astronomique de l'Universit{\'e} de Gen{\`e}ve, Chemin Pegasi 51, 1290 Versoix, Switzerland
\label{geneva}
\and Instituto de Astrof\'isica de Andaluc\'ia (IAA-CSIC), Glorieta de la Astronom\'ia s/n, 18008 Granada, Spain 
\label{iaa}
\and
Department of Space, Earth and Environment, Astronomy and Plasma Physics, Chalmers University of Technology, 412 96 Gothenburg, Sweden
\label{chalmers}
\and
Instituto de Astrof\'isica e Ci\^encias do Espa\c{c}o, Universidade do Porto, CAUP, Rua das Estrelas, 4150-762 Porto, Portugal
\label{ia-porto}
\and
Aix-Marseille Univ., CNRS, CNES, LAM, Marseille, France
\label{marseille}
\and
NASA Ames Research Center, Moffett Field, CA 94035, USA
\label{ames}
\and
Instituto de Astrof\'isica de Canarias (IAC), 38205 La Laguna, Tenerife, Spain 
\label{iac}
\and
Departamento de Astrof\'isica, Universidad de La Laguna (ULL), 38206, La Laguna, Tenerife, Spain
\label{ull}
\and
Astrophysics Group, Keele University, Staffordshire ST5 5BG, U.K. 
\label{keele}
\and
Department of Astronomy, University of California Berkeley, Berkeley, CA 94720, USA
\label{berkeley}
\and
Department of Physics and Astronomy, Vanderbilt University, Nashville, TN 37235, USA
\label{nashville}
\and
Departamento de F\'isica e Astronomia, Faculdade de Ci\^encias, Universidade do Porto, Rua do Campo Alegre, 4169-007 Porto, Portugal
\label{uporto}
\and
Institute of Astronomy, University of Cambridge, Madingley Road, Cambridge, CB3 0HA, UK
\label{ioa}
\and
Chalmers University of Technology, Department of Space, Earth and Environment, Onsala Space Observatory, SE-439 92 Onsala, Sweden
\label{chalmers2}
\and
Department of Astronomy, University of Maryland, College Park, MD 20742, USA
\label{maryland}
\and
Center for Space and Habitability, University of Bern, Gesellschaftsstrasse 6, 3012, Bern, Switzerland
\label{csh}
\and
Carnegie Earth \& Planets Laboratory, 5241 Broach Branch Road NW, Washington, DC 20015, USA
\label{carnegie}
\and
Department of Astrophysical Sciences, Princeton University, Princeton, NJ 08544, USA
\label{princeton}
\and
SETI Institute/NASA Ames Research Center, Moffett Field, CA 94035, USA
\label{seti}
\and
The Observatories of the Carnegie Institution for Science, 813 Santa Barbara St, Pasadena, CA 91101, USA
\label{carnegie-ca}
\and
International Center for Advanced Studies (ICAS) and ICIFI (CONICET), ECyT-UNSAM, Campus Miguelete, 25 de Mayo y Francia, (1650) Buenos Aires, Argentina.
\label{icas}
\and
Department of Astronomy of the University of Geneva, 51 Chemin de Pegasi, 1290 Versoix, Switzerland
\label{geneva-inst}
\and
Th\"uringer Landessternwarte Tautenburg, Sternwarte 5, 07778, Tautenburg, Germany
\label{tautenburg}
\and
Space Telescope Science Institute, 3700 San Martin Drive, Baltimore, MD, 21218, USA
\label{stsci}
\and
Max Planck Institute for Astronomy, K\"onigstuhl 17, 69117 Heidelberg, Germany
\label{mpia}
\and
Department of Physics and Astronomy, University of New Mexico, 210 Yale Blvd NE, Albuquerque, NM 87106, USA
\label{newmex}
\and
Department of Physics and Kavli Institute for Astrophysics and Space Research, Massachusetts Institute of Technology, Cambridge, MA 02139, USA
\label{mit}\\
\and
Center for Data Intensive and Time Domain Astronomy, Department of Physics and Astronomy, Michigan State University, East Lansing, MI 48824, USA
\label{michigan}
\and
Department of Earth, Atmospheric and Planetary Sciences, Massachusetts Institute of Technology, Cambridge, MA 02139, USA
\label{mit-earth}
\and
Department of Aeronautics and Astronautics, MIT, 77 Massachusetts Avenue, Cambridge, MA 02139, USA
\label{mit-aero}
\and
Facultad de Ciencias Astron\'omicas y Geof\'isicas, Universidad Nacional de La Plata, Paseo del Bosque s/n, (B1900) Buenos Aires, Argentina.
\label{laplata}
\and
Department of Astronomy, Tsinghua University, Beijing 100084, People's Republic of China
\label{tsinghua}
\and
University of Southern Queensland, Centre for Astrophysics, West Street, Toowoomba, QLD 4350 Australia
\label{queensland}
}

   \date{Received XXX; accepted YYY}

 
  \abstract
   {The current architecture of a given multi-planetary system is a key fingerprint of its past formation and dynamical evolution history. Long-term follow-up observations are key to complete their picture.}
   {In this paper we focus on the confirmation and characterization of the components of the \toi{} planetary system, where TESS detected a Neptune-size planet candidate in a very close-in orbit around a late K-dwarf star.}
   {We use a set of precise radial velocity observations from HARPS, PFS and CORALIE instruments covering more than two years in combination with the TESS photometric light curve and other ground-based follow-up observations to confirm and characterize the components of this planetary system.}
   {We find that \toib{} is a transiting close-in ($P_b\sim 1.82$~days) mini-Neptune planet ($m_b=9.1^{+1.1}_{-1.0}$~\Mearth{}, $R_b=2.765^{+0.088}_{-0.097}$~\Rearth{}), thus placing it on the {lower boundary} of the hot-Neptune desert ($T_{\rm eq,b}=941\pm31$~K). The analysis of its internal structure shows that \toib{} is a volatile-rich planet, suggesting it underwent an inward migration. The radial velocity model also favors the presence of a second massive body in the system, \toic{}, with a long period of $P_c=1700^{+290}_{-280}$~days and a minimum mass of $m_{c}\sin{i_c}=11.3^{+1.1}_{-0.9}$~\Mjup{}, and with a highly-eccentric orbit of $e_c=0.628^{+0.043}_{-0.036}$.}
   {The \toi{} planetary system is one of the few around K-dwarfs known to have this extended configuration going from a very close-in planet to a wide-separation gaseous giant. \toib{} has a transmission spectroscopy metric of 93, and it orbits a moderately bright ($G=11.3$ mag) star, thus becoming an excellent target for atmospheric studies. The architecture of this planetary system can also provide valuable information about migration and formation of planetary systems.}


	\titlerunning{TOI-969}
	\authorrunning{Lillo-Box et al.}

   \maketitle
%

\section{Introduction}
So far, 860 multi-planetary systems (with two or more planets) around stars other than the Sun have been found hosting a total of 2050 planets according to the NASA Exoplanet archive \citep{akeson13} to date. In contrast to our Solar System, the components of most of these known multi-planet systems reside in close-in compact configurations \citep[e.g.,][]{gillon16}, mainly biased by the observational techniques. This is especially relevant in the case of late-type stars, with effective temperatures below 4600 K, for which 90\% of multi-planetary systems have all their detected planets interior to the snow line (the distance from the parent star within which volatiles can condense into ice grains), and 91\% of the known long-period planets are massive and isolated \citep{lillo-box22}. Only a small number of systems in this stellar regime show components on both sides of the snow line (e.g., WASP-107 by \citealt{anderson17,piaulet20}, or GJ\,1148 by \citealt{trifonov18}). This is a clear fingerprint of the planet migration and formation history. Thus finding new systems covering these regimes is an important step towards comprehending planetary histories.

Exoplanet exploration has also revealed different exoplanet populations, including zones in the parameter space where there is a clear dearth of extrasolar planets. One of the most intriguing is the hot-Neptune desert \citep{szabo11,owen18}, a deficit of icy to sub-Jovian planets ($R_p \sim 2-10$\Rearth) at close-in orbits ($P<5$ days). Several theories have been proposed to explain the emptiness of this region, such as atmospheric mass loss of low-mass gaseous giants unable to retain their atmospheres due to the strong stellar irradiation suffered along their inward migration \citep{lopez14,owen13}; a consequence of the core accretion scenario in forming gas giants more preferentially than Neptune-like planets \citep[e.g.,][]{ida08}; or a faster migration of the lowest-mass end of gas giants preventing them to accrete large amounts of gas \citep{flock19}.

Interestingly, both the Kepler \citep{borucki10} and TESS (Transiting Exoplanet Survey Satellite, \citealt{ricker14}) missions, assisted by ground-based follow-up observations, have sparsely populated this desert with planets displaying unusual properties. Two clear examples of this are TOI-849\,b \citep{armstrong20} and LTT\,9779\,b \citep{jenkins20}, two ultra-short period planets in the middle of the hot-Neptune desert. The advantage of the planets detected by TESS in this regime is that they usually orbit stars much brighter than those detected by Kepler. This, combined with the transiting nature of these detections and the precise masses achievable due to their proximity to their host star, opens the possibility to study their interior properties (through internal structure analysis, e.g., \citealt{delrez21}) and composition with the James Webb Space Telescope (JWST, \citealt{gardner06}).

In this paper, we present the confirmation and characterization of the planetary system \toi{} (also known as TIC\,280437559, TYC\,183-755-1, Gaia\,DR3\,3087206553745290624), composed of at least two very distinct companions and whose architecture and properties of the individual components may be a test bench to study theories of planet formation and migration in the cool regime of solar-like stars. The inner component, a short-period mini-Neptune, was detected by the TESS mission. The outer component is an eccentric massive body in the boundary between planet and brown dwarf domain orbiting with a long orbital period beyond the snow line that we report here for the first time. In Sect.~\ref{sec:observations}, we describe the observations from different instruments and the techniques that have been used to characterize the system. In Sect.~\ref{sec:StePar}, we use these observations to constrain the stellar parameters and in Sect.~\ref{sec:Analysis} to unveil the properties of the planetary system. We discuss the results of this analysis in Sect.~\ref{sec:Discussion}, and we finish in Sect.~\ref{sec:Conclusions} by providing our final conclusions.

\section{Observations}
\label{sec:observations}

\subsection{High-precision photometry}
\label{sec:Obs.Phot}

\subsubsection{TESS}
\label{sec:tess}

Observations of \toi{} are available in sectors 7 (from Cycle 1) and 34 (Cycle 3), in both cases with camera \#1. In the case of Sector 7 the cadence of the time series correspond to 1800\,s while in the case of {Sector 34} a cadence of 600\,s is available. We downloaded the light curve files from the MAST archive\footnote{\url{https://archive.stsci.edu}} corresponding to the data extraction by the TESS-SPOC pipeline (Caldwell et al 2020), which uses the same codebase as the Science Processing Operations Center (SPOC; \citealt{jenkins16}). We use the Presearch Data Conditioned Simple Aperture Photometry (PDCSAP; \citealt{stumpe12,stumpe14,smith12}) flux and remove any data points marked as outliers or having non-numeric values. 

\begin{figure}
\centering
\includegraphics[width=0.5\textwidth]{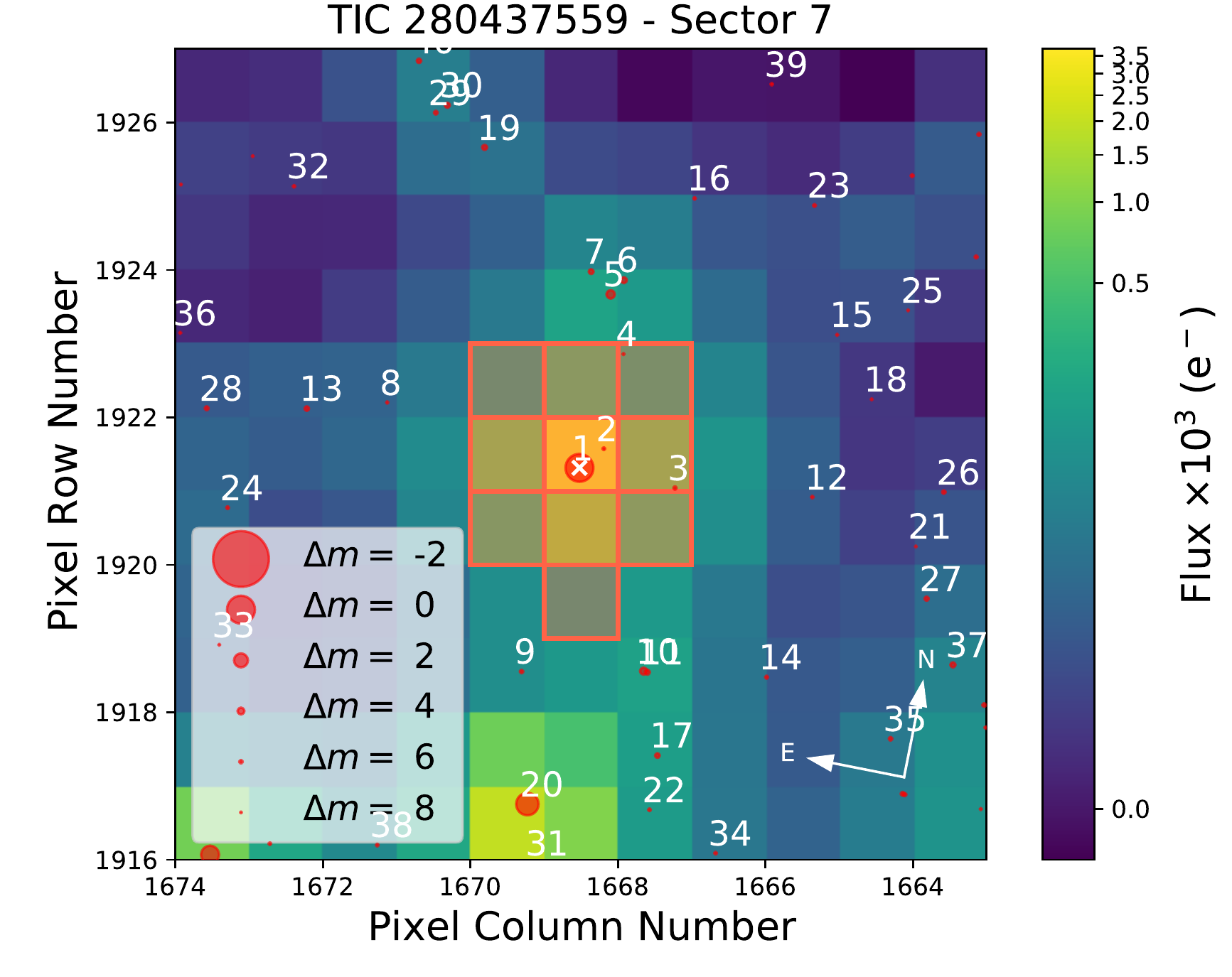}
\includegraphics[width=0.5\textwidth]{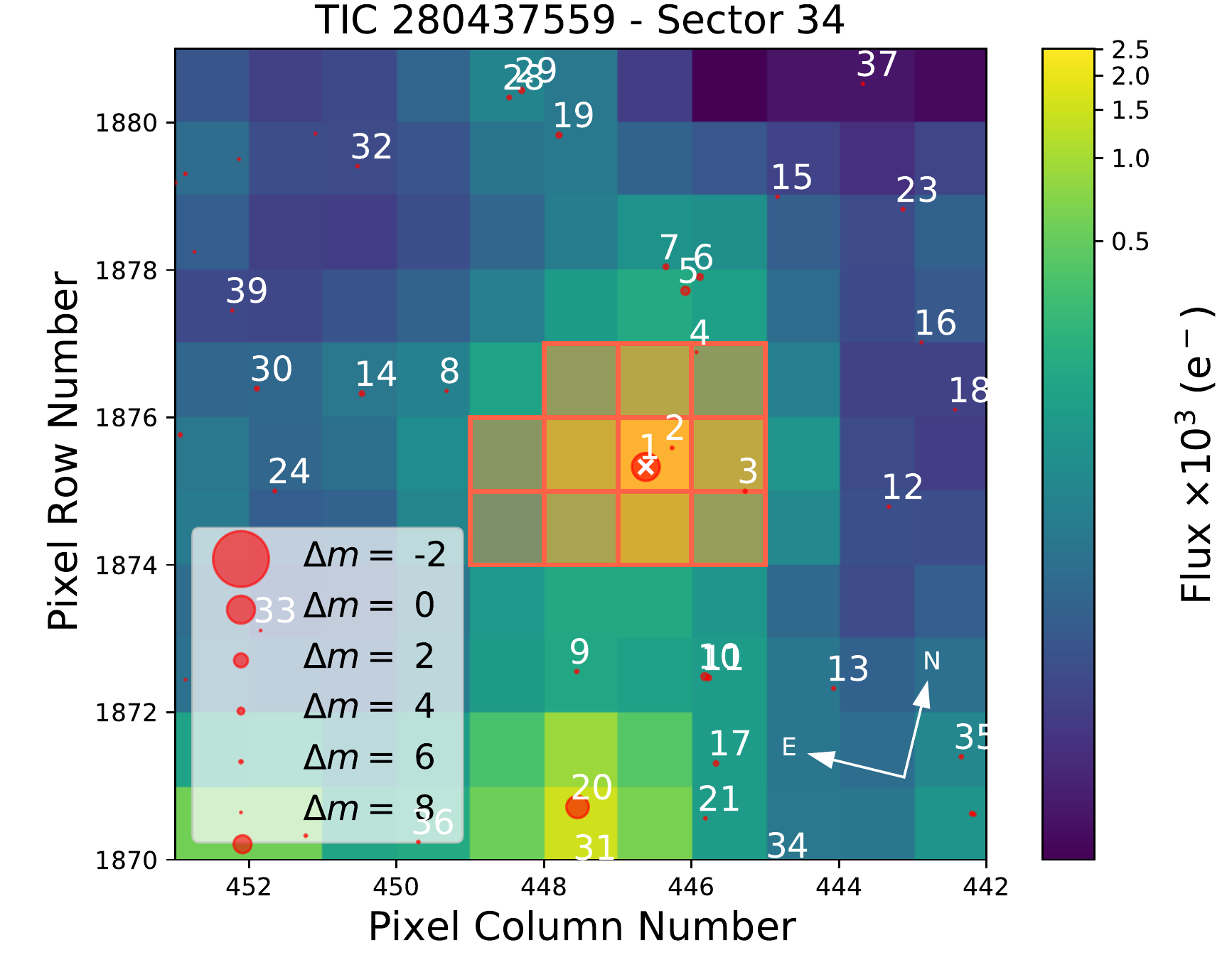}
\caption{Target Pixel File (TPF) corresponding to \toi{} observations by the TESS mission in sectors 7 (upper panel) and 34 (lower panel). In both panels, the sources from the Gaia catalog DR3 are shown as red circles, with their size corresponding to the magnitude contrast against \toi{} (marked with the label "1" and a white cross). The aperture used by the TESS-SPOC pipeline is shown as a shaded red region on each panel. The figures are computed using the \texttt{tpfplotter} algorithm by \cite{aller20}. }
\label{fig:TPF}
\end{figure}

We first inspect the target pixel files (TPF) for each sector by using the \texttt{tpfplotter} algorithm\footnote{This code is publicly available in Github through the following URL:  \url{https://github.com/jlillo/tpfplotter}.} \citep{aller20} to check for possible contaminant sources within the pipeline aperture. In Fig.~\ref{fig:TPF} we show the average TPF for each sector. As shown, both sectors include three additional sources in the aperture. However, their magnitude contrast in the Gaia passband are 6.6 mag (source labelled as \#2 in Fig.~\ref{fig:TPF}), 6.1 mag (source \#3) and 7.6 mag (source \#4). Overall, the contamination coming from these three sources is well below 0.1\%. Consequently, we assume the flux as originating only from \toi{}.

\begin{figure*}
\centering
\includegraphics[width=1\textwidth]{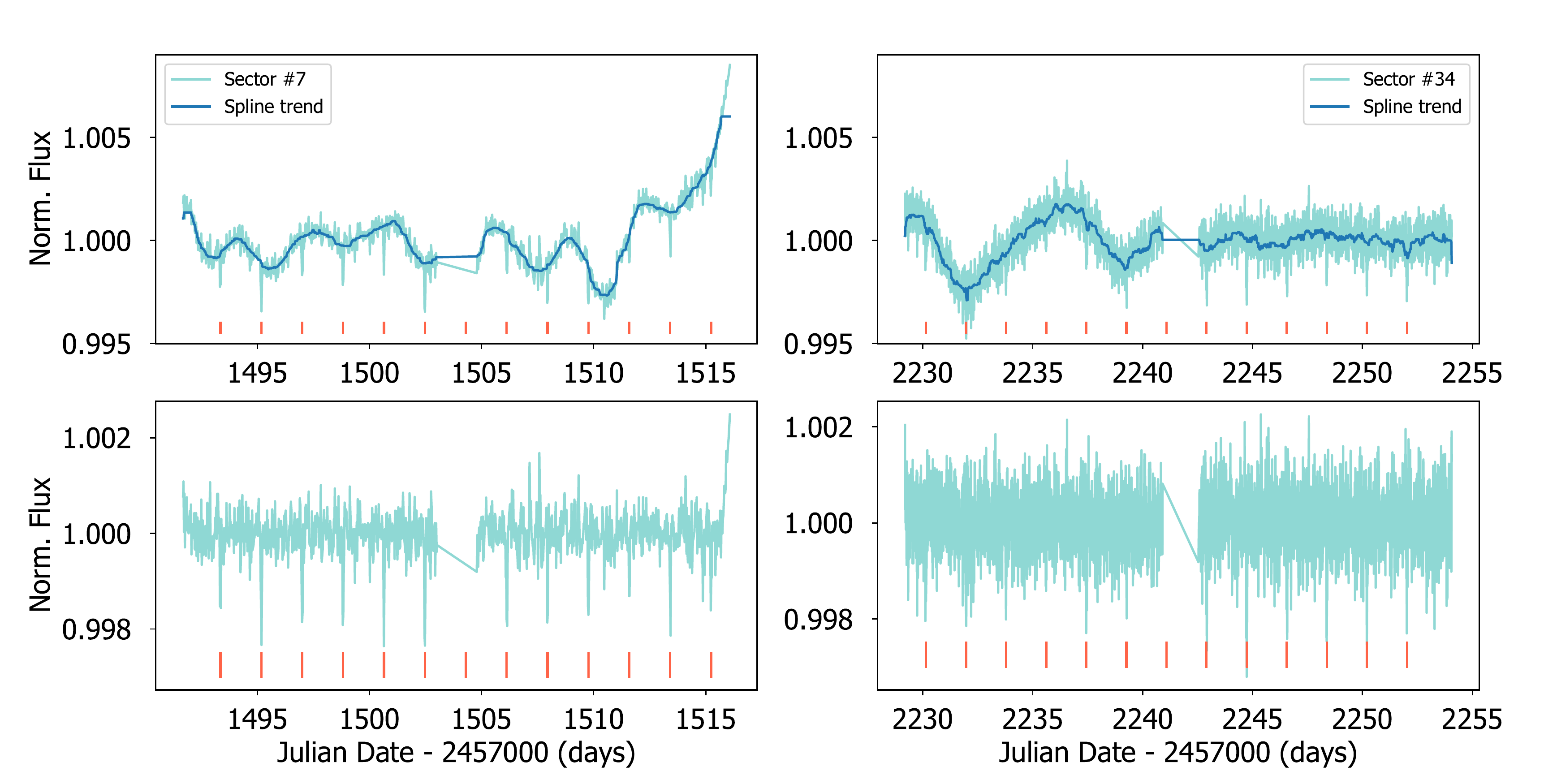}
\caption{TESS photometric timeseries from sectors \#7 (left panels) and \#34 (right panels). Upper panels display the PDCSAP flux and a simple median filter compted trend (blue solid line). The detrended light curves are shown in the lower panels and the transits from \toib{} are marked by {vertical red lines at the bottom of each panel.} }
\label{fig:TESS}
\end{figure*}

The photometric time series for both sectors are presented in Fig.~\ref{fig:TESS} and Table~\ref{tab:TESSdata}, together with their corresponding Lomb-Scargle periodograms. They show clear large photometric variations that can be attributed to the stellar activity, with the main periodicities around 8 and 12 days. These might be the harmonics of the possible rotation period at 24 days, as we see in the spectroscopic activity indicators (see Sect.~\ref{sec:Obs.Spec}). Besides, the light curve clearly shows the dips corresponding to the transits of the 1.8-day period planet candidate alerted by the TESS Science Office (see vertical {line marks} in Fig.~\ref{fig:TESS}).

Based on this alert, we observed the transit of the planet candidate from different ground-based facilities.\\ 

\subsubsection{LCOGT-MuSCAT3}
\label{sec:lco}

A full transit of \toib{} was observed simultaneously in Sloan $g'$, $r'$, $i'$, and Pan-STARRS $z$-short bands on UTC 2021 January 08 using the Las Cumbres Observatory Global Telescope \citep[LCOGT;][]{Brown:2013} 2\,m Faulkes Telescope North at Haleakala Observatory on Maui, Hawai'i. The telescope is equipped with the MuSCAT3 multi-band imager \citep{Narita:2020}. We used the {\tt TESS Transit Finder}, which is a customized version of the {\tt Tapir} software package \citep{Jensen:2013}, to schedule our transit observations. The images were calibrated using the standard LCOGT BANZAI pipeline, and photometric data were extracted using {\tt AstroImageJ}. The images were mildly defocused and have typical stellar point-spread-functions with FWHM of $\sim 1\farcs6$, and circular apertures with radius $4\arcsec$ were used to extract the differential photometry. The apertures exclude essentially all of the flux from the nearest Gaia DR3 neighbor (TIC 280437561) $9\arcsec$ northwest of the target (labelled as \#2 in Fig.~\ref{fig:TPF}). The transit was detected on target in all four filter bands. The final light curves are presented in Table~\ref{tab:GBdata}.

\subsubsection{MEarth}
\label{sec:MEarth}

One transit of \toib{} was observed using the MEarth-South telescope array \citep{irwin15} at Cerro Tololo Inter-American Observatory (CTIO), Chile on UT 2020 February 4. Six telescopes were used to obtain photometry of the target star, defocused to half flux diameter (HFD) of 6 pixels, equivalent to 5 arcsec given the pixel scale of 0.84 arcsec/pix.  Exposure times were 60 seconds.  One telescope was used to obtain photometry of any contaminating fainter sources at higher angular resolution, with the target saturated, using the same exposure time.  Observations were gathered starting at evening twilight and continuing until the target set below an airmass of 2.

Data were reduced using the standard MEarth processing pipeline (e.g. \citealt{irwin07,berta12}) with a photometric extraction aperture of $r = 10$ pixels (8.4 arcsec). The light curves contain meridian flips during transit.  These were taken into account in the data reduction by using separate magnitude zero points for each combination of telescope and meridian side to remove residual flat fielding error. The final extracted photometric time series is presented in Table~\ref{tab:GBdata}.

\subsubsection{WASP}
\label{sec:wasp}
The field of \toi{} was observed between 2009 and 2012 by the WASP (Wide Angle Search for Planets) transit-search survey \citep{wasp}. A total of 14\,600 photometric data points were obtained, observing on clear nights when the field was visible with a typical cadence of 15 mins. The 200-mm, f/1.8 Canon lenses were backed by 2kx2k CCDs. \toi{} is five magnitudes brighter than any other star in the 48-arcsec extraction aperture. Table~\ref{tab:WASPdata} presents the photometric time series. Following the methods described in \citet{maxted11}, the Lomb-Scargle periodogram of this long-term dataset reveals a signal at a period of 24.6 $\pm$ 0.6 d, together with power at the 12.3-d first harmonic. The modulation has an amplitude of 8 mmag and has a false-alarm likelihood below 10$^{-3}$. This signal is similar to that described above in the TESS data, and its origin clearly points to the rotational periodicity of the star {(see Sect.~\ref{sec:StePar} for a detailed discussion about the stellar rotation period)}.

\subsection{High-spatial resolution imaging}
\label{sec:Obs.HRim}

The presence of close stellar companions to a star hosting a planet candidate (either physically bound of chance aligned) can create false-positive exoplanet detections under specific configurations (e.g., eclipsing binaries) or provide additional flux leading to an underestimated planetary radius and incorrect exoplanet parameters (see, e.g., \citealt{lillo-box12,lillo-box14b,ciardi15,furlan17}). Besides the large-separation sources described in the previous section through the analysis of the Gaia EDR3 catalog (\citealt{gaia21}), we explored the close-in region ($<3$ arcsec) around the candidate host star to unveil previously unresolved companions through high-spatial resolution images at different wavelengths.

\begin{figure}
\centering
\includegraphics[width=0.5\textwidth]{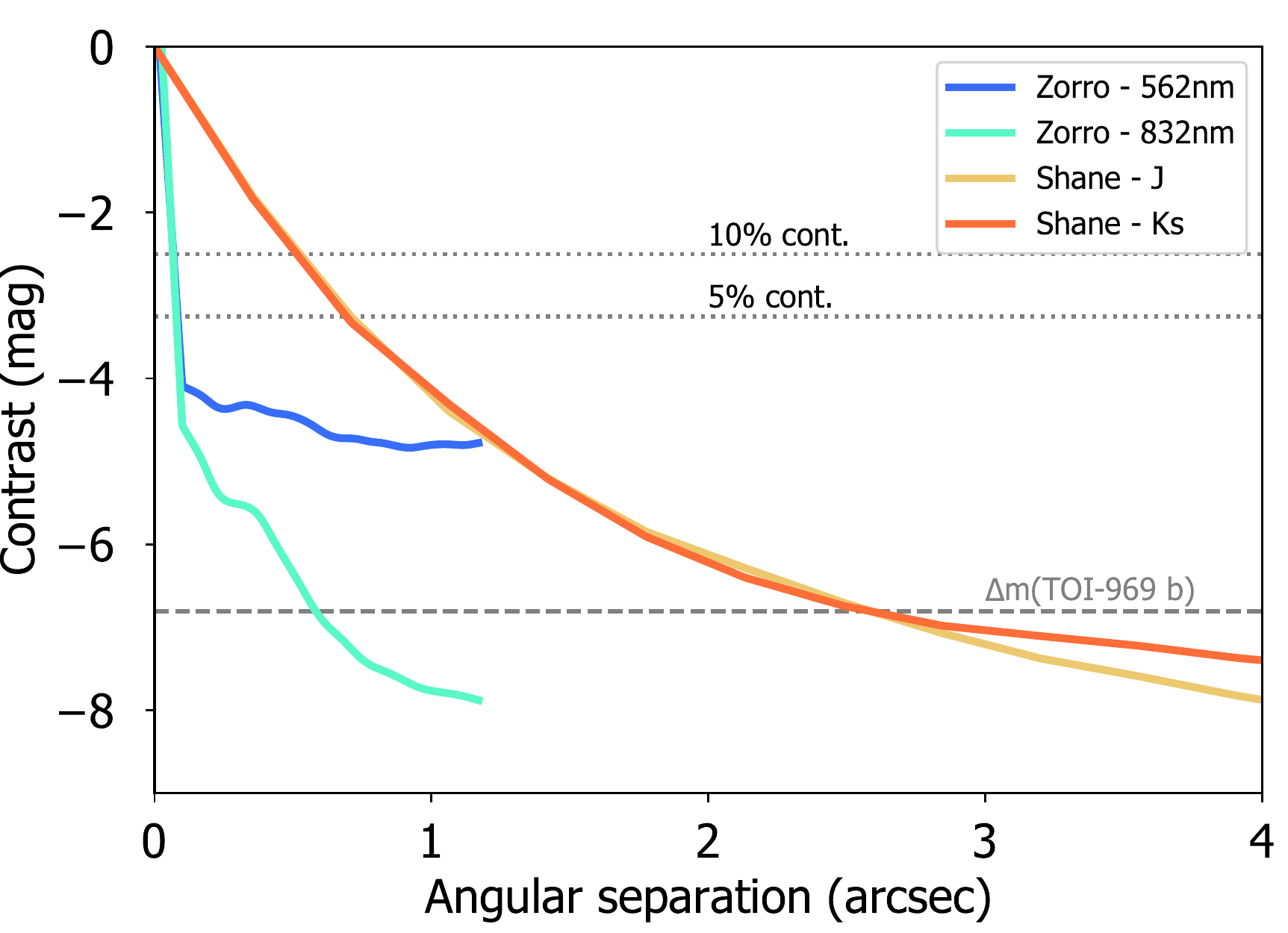}
\caption{{High-spatial resolution contrast curves for \toi{} from the Zorro speckle instrument on Gemini South (blue curve for the 562\,nm band and green curve for the 832\,nm band), and the 3m Shane telescope in the J-band (orange) and Ks-band (red). The two dotted lines correspond to 10\% (upper) and 5\% (lower) contamination levels in the lightcurve. The dashed line corresponds to the maximum contrast that a blended binary could have to mimic the transit depth of \toib{}.}}
\label{fig:HRimaging}
\end{figure}

\toi{} was observed on 2020 January 14 UT using the Zorro speckle instrument on Gemini South \citep{scott21}. Zorro provides simultaneous speckle imaging in two bands (562\,nm and 832\,nm) with output data products including a reconstructed image and robust contrast limits on companion detections \citep{howell11,howell16}. The night was clear, had a slight breeze, and good seeing (0.6 to 1.0 arcsec) during the observations. Figure~\ref{fig:HRimaging} (left panel) shows our limiting magnitude contrast curves and our 832\,nm reconstructed speckle image. We find that \toi{} is a single star with no companion brighter than 5 to 8 magnitudes from the diffraction limit out to 1.2 arcsec. At the distance to \toi{} (77~pc, see Sect.~\ref{sec:StePar}) these angular limits correspond to spatial limits of 1.5 to 92 AU.

We also observed \toi{} on UT 2019 October 15 using the ShARCS camera on the Shane 3-meter telescope at Lick Observatory \citep{2012SPIE.8447E..3GK, 2014SPIE.9148E..05G, 2014SPIE.9148E..3AM}. Observations were taken with the Shane adaptive optics system in natural guide star mode in order to search for nearby, unresolved stellar companions. We collected two sequences of observations, one with a K$_{\rm S}$ filter ($\lambda_0 = 2.150$ $\mu$m, $\Delta \lambda = 0.320$ $\mu$m) and one with a $J$ filter ($\lambda_0 = 1.238$ $\mu$m, $\Delta \lambda = 0.271$ $\mu$m), and reduced the data using the publicly available \texttt{SImMER} pipeline \citep{2020AJ....160..287S}\footnote{\url{https://github.com/arjunsavel/SImMER}}. Our reduced images and corresponding contrast curves are shown in Fig.~\ref{fig:HRimaging} (central and right panels). We find no nearby stellar companions within our detection limits.

\subsection{High-resolution spectroscopy}
\label{sec:Obs.Spec}

\subsubsection{HARPS}
\toi{} was observed intensively with the HARPS spectrograph (\citealt{mayor03}) on the ESO 3.6\,m telescope at La Silla Observatory, Chile, from 2020 November 10 to 2021 March 24. In total, 66 spectra were obtained under the programmes 1102.C-0249 (PI: Armstrong), and 106.21TJ.001 (PI: Gandolfi). HARPS is a stabilized high-resolution spectrograph with a resolving power of R$\sim$115\,000, capable of sub-\ms{} RV precision. We used the instrument in high-accuracy mode with a 1 arcsec science fibre on the star and a second fibre on sky to monitor the sky-background during exposure

\begin{figure*}
\centering
\includegraphics[width=1\textwidth]{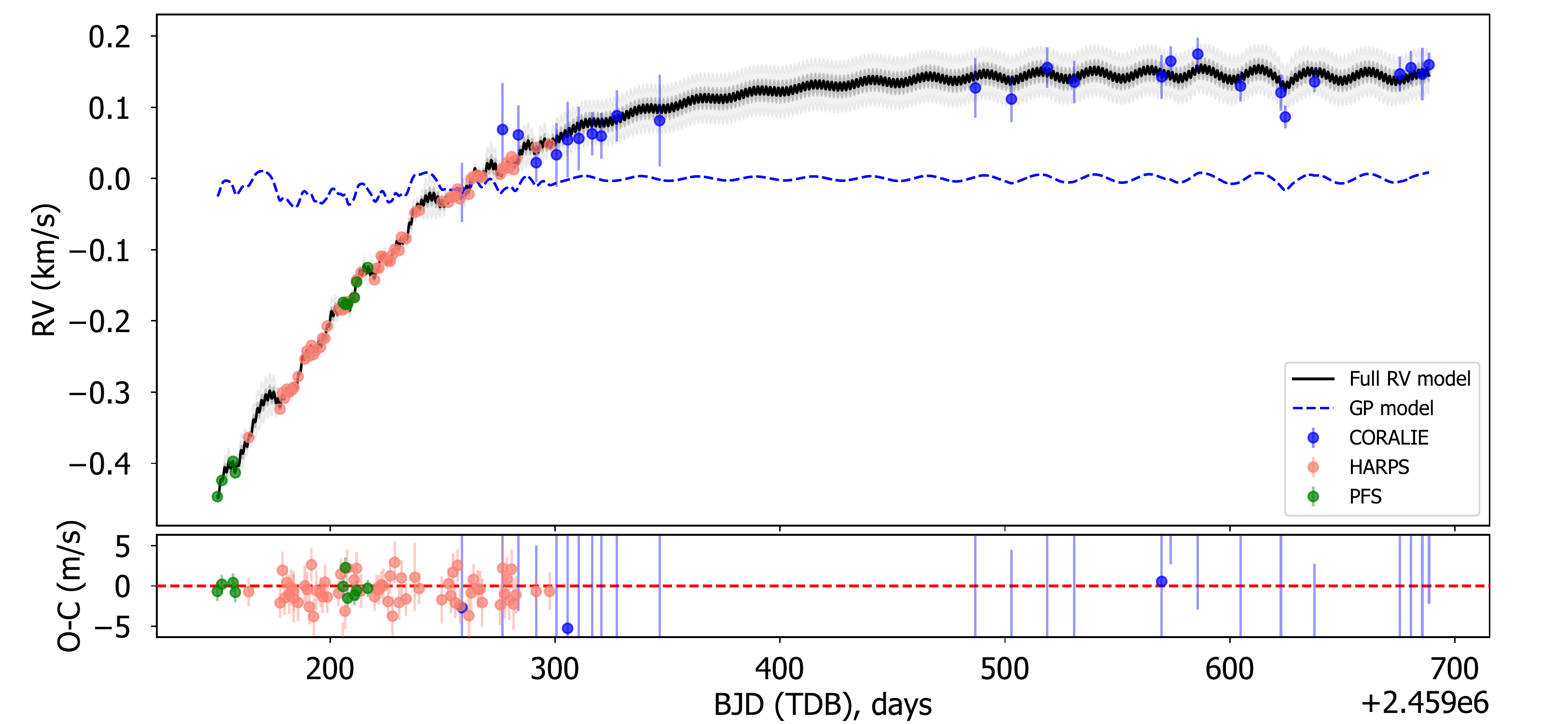}
\caption{Radial velocity time series and model inference from the joint radial velocity and light curve analysis in Sect.~\ref{sec:joint}, including the GP noise (blue dashed line) and full RV model (black solid line for the median posterior model and gray shaded regions for 68.7\% - in dark gray- and 99.7\% - in light gray- confidence intervals). }
\label{fig:RVresult}
\end{figure*}

\begin{figure}
\centering
\includegraphics[width=0.5\textwidth]{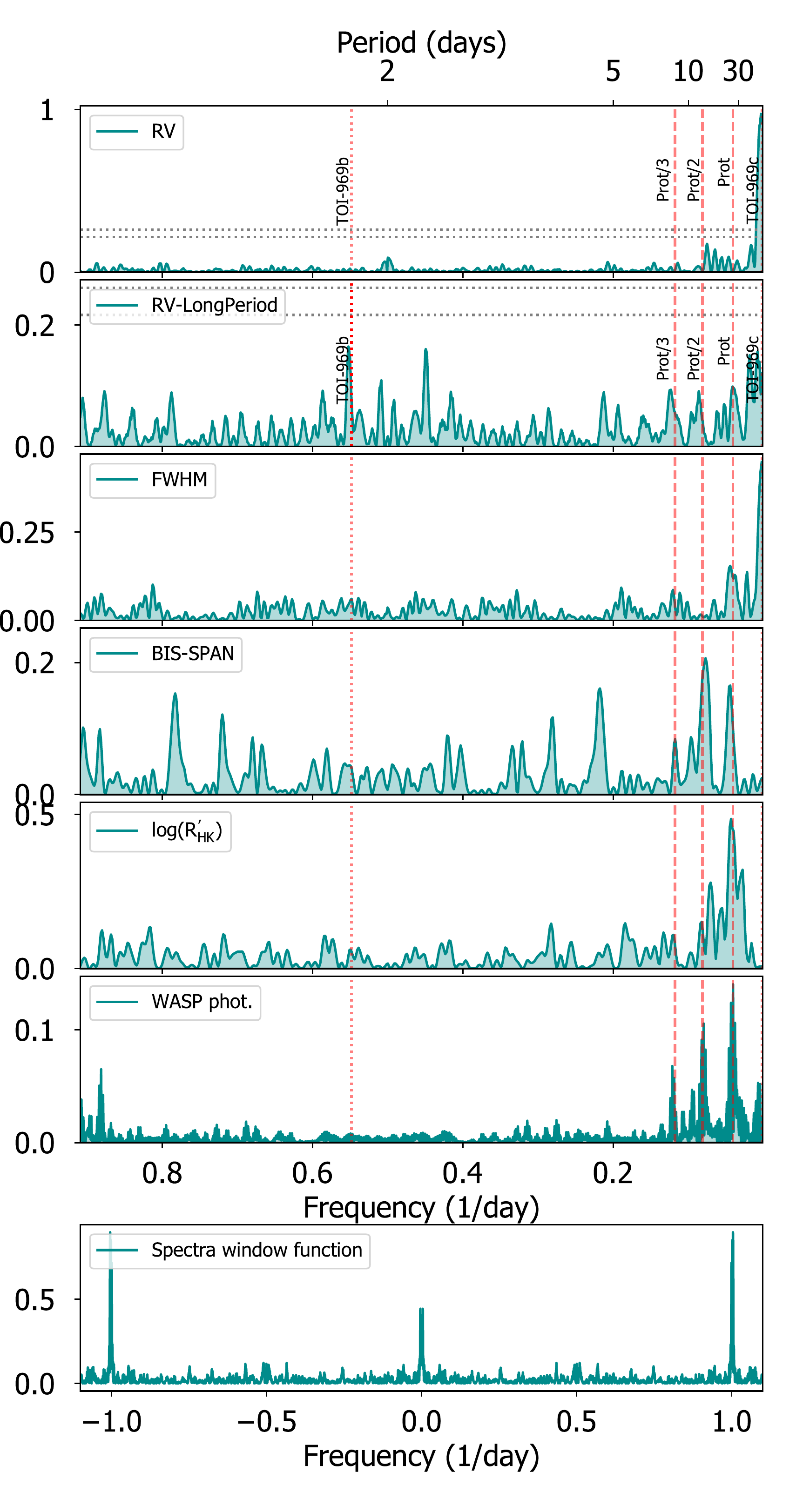}
\caption{Periodogram of the radial velocity and activity indicators time series of the HARPS dataset. Panel (a) corresponds to the original RV time series, while panel (b) corresponds to the time series after a simple linear detrending. In panel (c) we include an additional linear detrending corresponding to the activity indicator most correlated with the RVs (in this case the BIS-span). Panels (d)-(e) include the periodogram of the full-width at half maximum (FWHM) of the cross-correlation function and the bisector span, respectively. Finally, in the bottom panel, we show the window function corresponding to the HARPS sampling. The periodicity corresponding to \toib{} and \toic{} are shown as vertical dotted lines in all panels. The stellar rotation period and its first two harmonics are shown as dashed vertical lines. }
\label{fig:RVperiodogram}
\end{figure}

RVs were determined with the standard (online) HARPS data reduction pipeline using a K5 binary mask for the cross-correlation (\citealt{pepe02}), and a K5 template for flux correction to match the slope of the spectra across Echelle orders. With a typical signal-to-noise ratio (SNR) per pixel of 30, we achieved an RV precision of 2.5~\ms{}. For each epoch the bisector span (BIS), contrast, and full width at half-maximum (FWHM) of the CCF were calculated, as well as the chromospheric activity indicators CaII H\&K, H$\alpha$, and Na. These RVs and the activity indicators are presented in Table~\ref{tab:RVdata}. We clearly detect an RV signal for \toib{}, as well as an additional long term trend (see Fig.~\ref{fig:RVresult}). When removing the latter signal from a simple Keplerian fit (a linear or quadratic fit is not enough to remove the long-term variability), the Lomb-Scargle periodogram of the HARPS measurements (see Fig.~\ref{fig:RVperiodogram}) shows three peaks at around 8, 12, and 24 days, compatible to those seen in the WASP photometric data (see Sect.~\ref{sec:wasp} {and the the periodogram in the WASP panel of Fig.~\ref{fig:RVperiodogram}}). Indeed, these periodicities appear as well in the activity indicators corresponding to the full-width at half-maximum of the CCF and the bisector span (third and fourth panels in Fig.~\ref{fig:RVperiodogram}, respectively). The signal of \toib{} is not statistically significant in the periodogram of this dataset (False Alarm Probability of FAP$>10$\%) alone but a clear peak at the 1.8 days periodicity clearly stands above other signals.  

\subsubsection{PFS}
We observed \toi{} with the Planet Finder Spectrograph \citep[PFS;][]{Crane:2006,Crane:2008,Crane:2010}, which is mounted on the 6.5 m Magellan II (Clay) Telescope at Las Campanas Observatory in Chile. PFS is a slit-fed echelle spectrograph with a wavelength coverage of $3910$-$7340$\ \AA. We used a 0.3\arcsec\ slit and $3 \times 3$\ binning, which yields a resolving power of $R \approx 110,000$. Wavelength calibration is achieved via an iodine gas cell, which also allows characterization of the instrumental profile. We obtained 10 spectra, observed through iodine, between UT 2020-Oct-27 and 2021-Jan-02. Exposure times ranged from 20 to 30 minutes. We also obtained an iodine-free template observation with a 90-minute exposure time. The radial velocities were extracted using a custom IDL pipeline following the prescriptions of \citet{Marcy:1992} and \citet{Butler:1996}, and achieved a mean internal precision of $1.2\ {\rm m\,s}^{-1}$. The velocities are presented in Table~\ref{tab:RVdata} and Fig.~\ref{fig:RVresult}.

\subsubsection{CORALIE}
TOI-969 was monitored by the CORALIE high-resolution echelle spectrograph mounted on the 1.2 m Euler telescope at La Silla Observatory~\citep{coralie}. 22 spectra were obtained between UT 2021-02-13 and 2022-04-15 each with SNR of 10-20, depending on sky-conditions and exposure time which was set between 1800 and 2700 seconds. The spectrum corresponding to the night of 2022-01-16 was discarded due to high contamination from the Moon. Radial velocity measurements were extracted by cross-correlating each spectra with a binary G2 mask \citep{baranne96}, using the standard CORALIE data-reduction pipeline. Given the relatively low SNR, a modest RV precision of 30-50~\ms{} was achieved. Line-shape diagnostics such as bisector-span and FWHM were derived for the cross-correlation function (CCF). 

The extracted radial velocities and associated uncertainties as well as the corresponding activity indicators obtained from the CCF analysis are presented in Table~\ref{tab:RVdata}.

\section{Stellar parameters}
\label{sec:StePar}

\toi{} is a moderately bright (V=11.6, \citealt{hog00}) star in the late-K dwarf regime. Table~\ref{tab:basic} summarises the main properties of this star. According to the Gaia (\citealt{gaia}) EDR3 data release \citep{gaiaDR3}, this star is located at a distance of $77.82\pm0.14$ pc (corresponding to a parallax of $12.849\pm0.023$ mas, \citealt{lindegren20}). Based on the \textit{Gaia} proper motions, we determine the projected galactic velocities of \toi{} (see Table~\ref{tab:basic}) and we conclude by using the relations from \cite{bensby03} that this star likely belongs to the galactic thin disk (with a probability $10^3$ times higher than belonging to the thick disk and more than $10^4$ times higher than belonging to the halo).

We derived the spectroscopic stellar parameters ($T_{\mathrm{eff}}$, $\log g$, microturbulence, [Fe/H]) by using the HARPS spectra and ARES+MOOG code, following the same methodology described in \cite{sousa14} and \cite{santos13}. The equivalent widths (EW) of iron lines were measured on the combined HARPS spectrum of \toi{} using the ARES code\footnote{The last version of ARES code (ARES v2) can be downloaded at \url{https://github.com/sousasag/ARES}} \citep{sousa07,sousa15}. The best set of spectroscopic parameters was found when we reach the ionization and excitation equilibrium. This process makes use of a grid of Kurucz model atmospheres \citep{Kurucz-93} and the radiative transfer code MOOG \citep{Sneden-73}. Since the star is cooler than 5200~K we used the appropriate iron linelist for our method presented in \citet[][]{Tsantaki-2013}. Following the same methodology as described in \citet[][]{sousa21}, we used the \textit{Gaia} EDR3 parallax and estimated the trigonometric surface gravity to be $\log{g}= 4.55 \pm 0.06$~dex. Stellar abundances of the elements were derived using the classical curve-of-growth analysis method assuming local thermodynamic equilibrium \citep[e.g.][]{adibekyan12,adibekyan15}. The same codes and models were used for the abundance determinations. {Finally, the mass, radius, and age of the star were derived using the \texttt{PARAM 1.3} web-interface\footnote{\url{http://stev.oapd.inaf.it/cgi-bin/param\_1.3}} \citep[][]{daSilva06} using our spectroscopic parameters as input values}. The results from this analysis are shown in Table~\ref{tab:StePar} (second column).

{We use the S-index provided by the HARPS DRS and calibrated to the Mount Wilson scale \citep{vaughan78} to determine the $\log{R^{\prime}_{HK}}$ for each of the individual spectra. We obtain an average value of $-4.381 \pm 0.002$ dex with a dispersion of 0.026 dex. This dispersion is significantly larger than the uncertainty of the individual measurements which points to certain level of activity. Indeed, in the periodogram of the $\log{R^{\prime}_{HK}}$ time series (see corresponding panel of Fig.~\ref{fig:RVperiodogram}), a clear peak at 24.6 days stands out, again pointing to this periodicity as the rotaton period of the star. We can use the average value of the $\log{R^{\prime}_{HK}}$ to estimate the rotation period of the star. Given the B-V color of this star being $1.09\pm0.33$ mag, we can apply the \cite{noyes84} activity-rotation relationships through the \texttt{pyrhk} code \citep{daSilva06} and using the \cite{middelkoop82} bolometric corrections. Following this, we obtain a rotation period of $9.9 \pm 3.2$ days. If we instead use the empirical relations from \cite{suarez-mascareno16} for K-dwarfs, we obtain a rotation period of $9.5 \pm 1.0$ days, compatible with the previous value. Hence, the activity-rotation empirical relations point to a period shorter than the one measured from the {periodogram of the} activity indicators {(see Sect.~\ref{sec:Obs.Spec} and panels 3-5 in Fig.~\ref{fig:RVperiodogram})} and the photometric time series from WASP {(see Sect.~\ref{sec:wasp} and panel 6 in Fig.~\ref{fig:RVperiodogram})}. As an additional test, we computed the \vsini{} of from the HARPS spectrum and obtained a value of 2.87 +/- 0.37 \kms{}. Assuming an inclination of 90$^{\circ}$, this corresponds to a rotation period for this star of $12.1 \pm 1.7$ days, again compatible within the uncertainties with the previous results from empirical relations. {For a rotation period of around 24 days, and assuming the same spin axis inclination of 90$^{\circ}$, we obtain a projected rotation velocity of around 1.3 \kms{}.}
We then warn that for very slow rotating stars (\vsini{}$< 3 $~\kms{}), measuring accurate \vsini{} values is a difficult task and our determination should be taken with care. Consequently, the rotation period of the star is unclear from the present data. In our analysis, we will assume the rotation period is $P_{\rm rot}=24.6 \pm 0.6$ days but we will account for the power at 12.3 days in the different indicators and empirical relations by using a rotation kernel which also accounts for $P_{\rm rot}/2$. }

{We can now use gyrochronology to estimate the age of the system. By applying the \cite{angus19} relations through the \texttt{stardate}\footnote{\url{https://github.com/RuthAngus/stardate}} code, we obtain an age of $2.03 \pm 0.17$~Gyr when assuming $P_{\rm rot}=24.6 \pm 0.6$ and the Gaia $B_p-R_p$ color of 1.42 mag \cite{gaia21}. If we use the 12.3 days rotation period obtained from the \vsini{}, we then obtain an age of 670~$\pm$~150~Myr. Given this young age estimation in the second case, we checked for the presence of lithium in the HARPS spectra and found none within the sensitivity limits. However, for this effective temperature it is expected that the Li has already been depleted at such age, so we cannot discard this young estimation based on the absence of lithium.}

\begin{figure}
\centering
\includegraphics[width=0.50\textwidth]{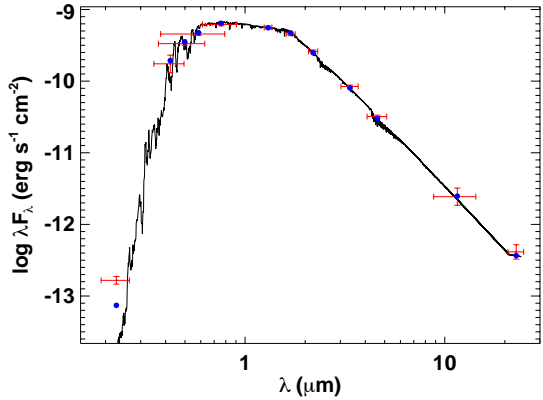}
\caption{Spectral energy distribution of TOI-969. Red symbols represent the observed photometric measurements, where the horizontal bars represent the effective width of the passband. Blue symbols are the model fluxes from the best-fit NextGen atmosphere model (black).}
\label{fig:sed}
\end{figure}

As an independent determination of the basic stellar parameters, we performed an analysis of the broadband spectral energy distribution (SED) of the star together with the {\it Gaia\/} EDR3 parallax \citep[with no systematic offset applied; see, e.g.,][]{stassun21}, in order to determine an empirical measurement of the stellar radius, following the procedures described in \cite{stassun16,stassun17,stassun18}. We pulled the $B_T V_T$ magnitudes from {\it Tycho-2}, the $JHK_S$ magnitudes from {\it 2MASS}, the W1--W4 magnitudes from {\it WISE}, the $G_{\rm BP} G_{\rm RP}$ magnitudes from {\it Gaia}, and the NUV magnitude from GALEX. Together, the available photometry spans the full stellar SED over the wavelength range 0.2--22~$\mu$m (see Figure~\ref{fig:sed}).

We performed a fit using NExtGen stellar atmosphere models, with the free parameters being the effective temperature ($T_{\rm eff}$) and metallicity ([Fe/H]). The remaining free parameter is the extinction $A_V$, which we fixed at zero due to the star's proximity. The resulting fit (Figure~\ref{fig:sed}) has a reduced $\chi^2$ of 1.1, excluding the GALEX NUV flux which indicates a moderate level of activity (see below). Integrating the (unreddened) model SED gives the bolometric flux at Earth, $F_{\rm bol} = 9.45 \pm 0.11 \times 10^{-10}$ erg~s$^{-1}$~cm$^{-2}$. Taking the $F_{\rm bol}$ and $T_{\rm eff}$ together with the {\it Gaia\/} parallax, we estimate the stellar radius; and using the empirical relations of \citet{torres10} and of \citet{Mann:2020} we estimate the stellar mass. All these values agree with spectroscopically derived parameters within the uncertainties and are reported in Table~\ref{tab:StePar} (third column).

\begin{table}[]
\setlength{\extrarowheight}{3pt}
\caption{\label{tab:basic} General and dynamical properties of  \toi{} obtained in this work.}
\begin{tabular}{lll}
 \hline 
Parameter & Value & Ref.$^{\dagger}$ \\ 
\hline 
IDs & \toi{}, TIC\,280437559, \\
	&  TYC\,0183-00755-1 \\
Gaia EDR3 ID & 3087206553745290624  & [1] \\
RA, DEC & 07:40:32.8, 02:05:54.92 &  [1]\\ 
Parallax (mas) & $12.849\pm0.023$    & [1]\\
Distance (pc) & $77.82\pm0.14$ & [1, 3] \\
$\mu_{\alpha}$ (mas/yr) & $-34.040\pm 0.025$  & [1]\\
$\mu_{\delta}$ (mas/yr) & $-77.949\pm 0.018$  & [1]\\
RV (km/s) & $-5.789 \pm 0.040$  &  [3] \\ 
G (mag) & 11.27  &  [1]\\
$B_p-R_p$ (mag) & 1.46  &  [1]\\
J (mag) & $9.596\pm0.026$ & [2] \\
Ks (mag) & $8.879\pm0.023$ & [2] \\
U (km/s) & 20.855 & [3], Sect.~\ref{sec:StePar}\\
V (km/s) & -2.101 & [3], Sect.~\ref{sec:StePar} \\
W (km/s) & -18.526 & [3], Sect.~\ref{sec:StePar}\\
Gal. population & Thin disk & [3], Sect.~\ref{sec:StePar} \\
\hline
\end{tabular}
\tablebib{
[1] \cite{gaia}; 
[2] \cite{cohen03}; 
[3] This work;
[4] \cite{mann15}.
}
\end{table}

\begin{table}[]
\setlength{\extrarowheight}{3pt}
\caption{\label{tab:StePar} Stellar physical properties of \toi{}.}
\begin{tabular}{lll} 
\hline 
Parameter & Spec.\tablefootmark{a} & SED\tablefootmark{b}  \\ 
\hline 
T$_{\rm eff}$ (K) & $4435\pm80$ & $4550\pm 75$ \\
$\log{\rm g}$ (dex)& $4.55\pm0.06$ & - \\
${\rm [Fe/H]}$ (dex) & $0.175\pm 0.019$ & - \\
M$_{\star}$ (\Msun{}) & $0.734 \pm 0.014$ & $0.730 \pm 0.037$ \\
R$_{\star}$ (\Rsun{}) & $0.671 \pm 0.015 $ & $0.681 \pm 0.023$ \\
$\log{R^{\prime}_{HK}}$ & $-4.381 \pm 0.026$ & - \\
Age (Gyr), \texttt{PARAM1.3} & $4.761 \pm 4.187$ & - \\
Age (Gyr), gyro  & $2.03 \pm 0.17$ & - \\
$\rm [Mg/H]$ (dex) & $0.17 \pm 0.15$ & - \\
$\rm [Si/H]$ (dex) & $0.17 \pm 0.13$ & - \\
\hline
\end{tabular}
\tablefoottext{a}{Parameters obtained from spectroscopic analysis.}
\tablefoottext{b}{Parameters obtained from analysis of the spectral energy distribution.}
\end{table}

\section{Analysis and results}
\label{sec:Analysis}
Due to the complexity of the system and the large amount of data in hand, we first perform an analysis to the individual datasets, namely the TESS light curve (Sect.~\ref{sec:TESSanalysis}) and to the radial velocity (Sect.~\ref{sec:RVanalysis}). From those independent analyses we obtain the relevant information about the models to be tested in the final joint (light curves and radial velocity) analysis in Sect.~\ref{sec:joint} as well as parameter ranges to appropriately set up the priors in this more complex and computationally expensive joint model.

\subsection{Light curve analysis}
\label{sec:TESSanalysis}

We first inspect the photometric data through an independent analysis of the TESS and LCOGT light curves. Our model is composed of a transit model assuming a single planet in the system (the one with the detected transits), defined by the period ($P_b$), a time of inferior conjunction ($T_{0,b}$), the orbital inclination ($i_b$), the planet radius ($R_{\rm b}$), and the stellar mass ($M_{\star}$) and radius ($R_{\star}$). The latter are used to check for possible phase curve variation effects like the ellipsoidal, Doppler beaming and reflection effects (which in turn are all negligible in this case). We use the limb-darkening parametrization described in \cite{kipping13} and apply the quadratic law with a pair of parameters per instrument and bandpass ($u_{\rm 1,j}$, $u_{\rm 2,j}$). Besides, we assume a photometric jitter and offset for each instrument ($\sigma_j$ and $F_{\rm 0,j}$, respectively). Finally, in the case of the ground-based observations, we use the pre-detrended time series and use a linear detrending with the airmass within our model by including a slope per bandpass for the LCOGT observations ($m_{\rm X,j}$). 

In order to account for the stellar variability seen in the TESS sectors, we also include a Gaussian Process (GP) with a {kernel composed of a mixture of harmonic oscillators designed to model stellar rotation and implemented in the \texttt{celerite2} \citep{celerite1,celerite2} package as the term named \texttt{RotationTerm}. One of the harmonic oscillators is devoted to the rotation period and the other one to half of the rotation period. This kernel includes five hyperparameters, namely an amplitude ($\eta_{\sigma,LC}$), a periodicity ($\eta_{\rho}$), a quality factor for the second oscillator ($\eta_{{\rm Q}_0}$), a difference between the two quality factors of both oscillators ($\eta_{\delta{\rm Q}}$), and a factor representing the relevance of one oscillator against the other ($\eta_{f}$).}

Based on this model including {35} parameters, we explore the parameter space informed by the observed time series by using the Monte-Carlo Markov Chain (MCMC) affine-invariant ensamble sampler \texttt{emcee} \citep{emcee}. We launch our MCMC with {140} walkers (four times the number of parameters) and a total 100\,000 steps per walker. {The priors for all parameters are detailed in Table~\ref{tab:LCalone}}. We use Gaussian priors for the orbital period and time of mid-transit as published by the TESS alert and using a broad standard deviation corresponding to five times the published uncertainty on these values. Gaussian priors are also used for the limb-darkening coefficients, with the mean value corresponding to the derived value using the \texttt{limb-darkening} code by \cite{espinoza15} using the ATLAS models and the stellar properties derived in Sect.~\ref{sec:StePar}; and a standard deviation of 0.1. The stellar mass and radius priors are also included as Gaussian distributions with the parameters described in Sect.~\ref{sec:StePar}. The priors for the rest of the parameters are set to uniform distributions within the ranges stated in Table~\ref{tab:LCalone}.

\begin{figure*}
\centering
\includegraphics[width=1\textwidth]{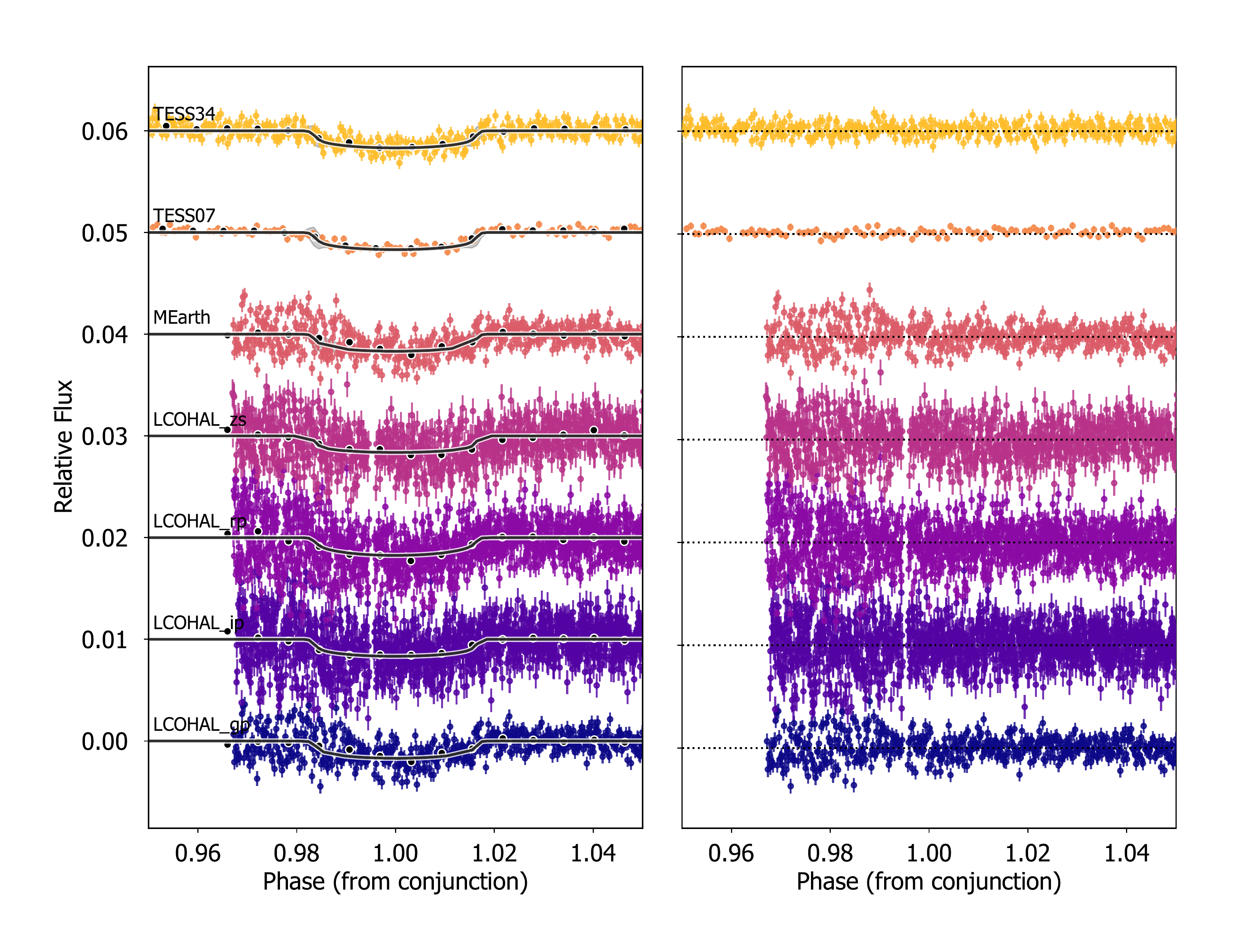}
\caption{Phase-folded light curves from \toi{} after removing the GP for the TESS data and the linear dependency for the LCOGT dataset. Left panels show the individual phase-folded light curves per instrument and the corresponding joint model. Black symbols correspond to 9-minute bins. Right panels show the residuals from the joint model.}
\label{fig:LCanalysis}
\end{figure*}

We perform a first burn-in phase with the previously mentioned setup. Subsequently, we use the maximum a posteriori set of parameters from this first phase to start a second chain initialised at those values and running for half the number of steps than in the first phase (i.e., 50\,000 steps per walker). The posterior distributions are computed by using the full chains from the second phase. The median and 68.7\% credible interval for each parameter are shown in Table~\ref{tab:LCalone}. 
The phase-folded light curves showing the median transit models from this analysis are shown in Fig.~\ref{fig:LCanalysis}.

\subsection{Radial velocity}
\label{sec:RVanalysis}

We now use the radial velocity dataset described in Sect.~\ref{sec:Obs.Spec} as well as the input information from the previous TESS analysis to scrutinize the Doppler information. 

To that end, we first build our model, which is composed of N Keplerians, each of them potentially including an orbital period ($P_i$), a time of inferior conjunction ($T_{0,i}$), a radial velocity semi-amplitude ($K_i$), and in the case of non-circular orbit models we also add eccentricity ($e_i$) and the corresponding argument of the periastron ($\omega_i$). Additionally, our models can include a slope ($m$) to account for possible linear trends (when this slope is assumed zero we include an "F" - for flat- in the model labels) and a quadratic trend (when this is added a "Q" is added to the model label). We also include a radial velocity offset per instrument ($\gamma_j$) and a jitter per instrument to account for all systematics not taken into account in the model ($\sigma_j$). In our analysis we test models that aim to account for the radial velocity variations caused by the stellar activity (potentially relevant as seen in the previous section in the TESS light curve) through Gaussian Process (GP) informed by an activity indicator. Based on the radial velocity periodogram presented in Sect.~\ref{sec:Obs.Spec}, the bisector span seems to match the radial velocity variations in a similar manner, showing strong periodicities at similar frequencies. This does not happen with the FWHM, which does not show significant variations other than a long-term periodicity. Hence, we use the BIS-span indicator to inform the GP about the activity-induced component of the radial velocity. Correspondingly, for each instrument, we add an offset ($\gamma_{\rm BIS,j}$) and a jitter ($\sigma_{\rm BIS,j}$).

We include a GP for the RVs and a GP for the BIS-span, both sharing all parameters except the amplitude of the kernel. We use the \texttt{celerite2} \citep{celerite1,celerite2} implementation to build the GPs and we choose the \texttt{RotationTerm} kernel (see Sect.~\ref{sec:TESSanalysis}) to account for the stellar activity. All hyper-parameters but the amplitude one ($\eta_{\sigma}$) are shared between the RV and BIS-span GPs.

{The prior distributions used in this analysis are summarised in Table~\ref{tab:RValone}.} Uninformative priors were generally used in most of the parameters involved in our model, except in some cases. We used Gaussian priors for the orbital period and mid-transit time for \toib{} centered at the values coming from the transit analysis from the previous section. This is because the precision in those parameters achievable with the transit method is higher than with RVs. We used log-uniform priors for the orbital period of the long-period candidate and the RV and BIS jitter, as well as for the amplitude hyper-parameters of the GP ($\eta_{\sigma ,FWHM}$ and $\eta_{\sigma}$). Finally, given the discussion presented in Sect.~\ref{sec:StePar}, we opted for a Gaussian prior on the hyper-parameter related to the rotation period of the star. 

The parameter space is explored by using \texttt{emcee}. We use a number of walker equal to four times the amount of free parameters and 100\,000 steps per walkers in a first burnout phase. The maximum a posteriori model from this first phase is selected as the initial guess for the final production, which now contains 100\,000 steps and the same number of walkers as the initial phase. The convergence of the chains is checked by estimating the autocorrelation time and the corresponding chain length, with the latest being at least 20 times longer than the autocorrelation time to consider convergence. We then use 15\% of the final flattened chain (typically composed of $10^5-10^6$ elements) to estimate the Bayesian evidence of each model ($\ln{\mathcal{Z}_i}$) and its corresponding uncertainty through the \texttt{perrakis} implementation\footnote{\url{https://github.com/exord/bayev}. A python implementation by R. D\'iaz of the formalism explained in \cite{perrakis14}.} \citep{diaz16}.

\begin{figure*}
\centering
\includegraphics[width=1\textwidth]{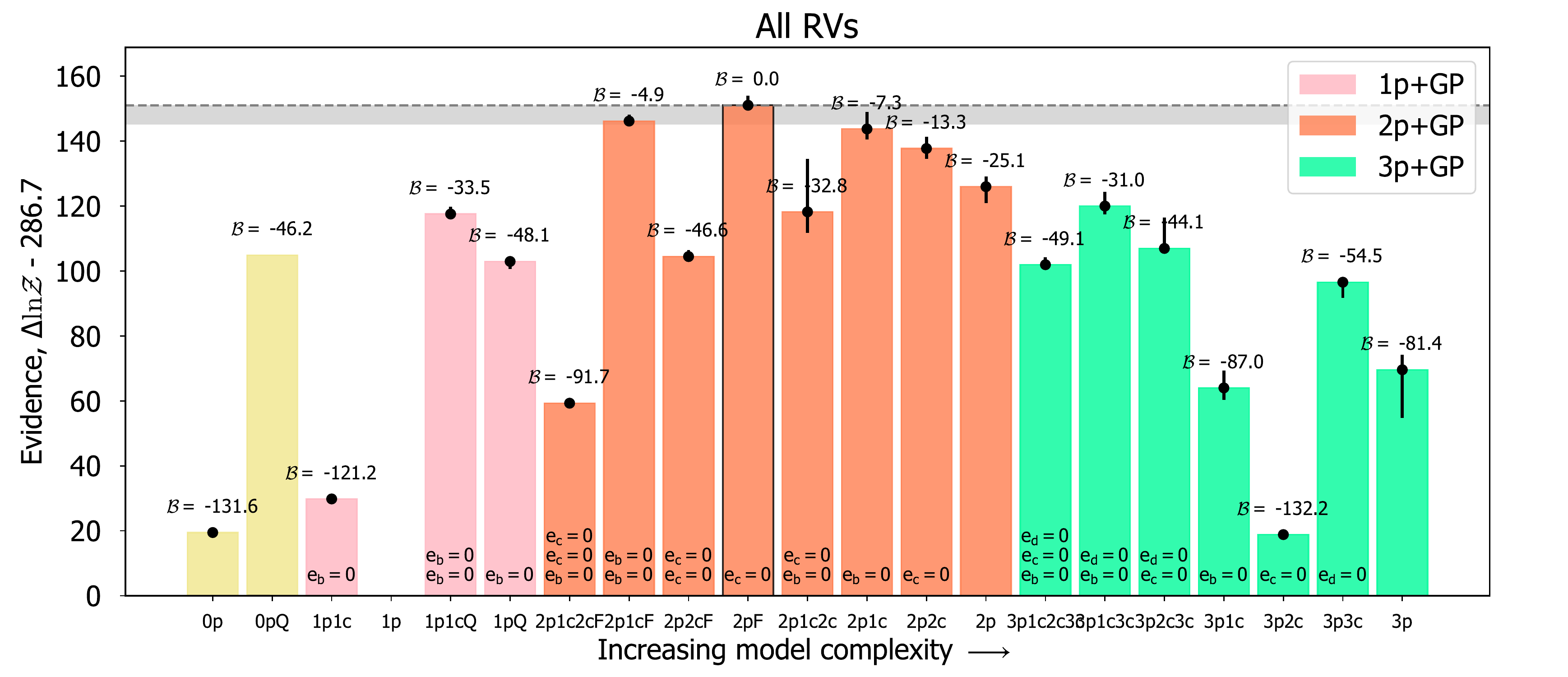}
\caption{Comparison of the logarithm of the Bayesian evidence ($\mathcal{B} \equiv \ln{\mathcal{Z}}$) for the different tested models (labelled in the X axis) in the case of the RV-only analysis (Sect.~\ref{sec:RVanalysis}). The label for the models correspond to Xp[P$_i$c][Z], where X is the number of planets assumed in the system, P$_i$c are the identifiers of the planets with assumed circular orbits, and Z indicates whether we are using a flat (F) slope (i.e., assuming slope$ = 0$) or a quadratic (Q) trend. The gray shaded region displays the region of $\Delta \mathcal{Z} = -6$ from the larger evidence model showing its strength against simpler models. }
\label{fig:RVmodelcomp}
\end{figure*}

 The preliminary analysis of the data already revealed the presence of different signals (see Sect.~\ref{sec:Obs.Spec}). Consequently, we test models with different number of planets, each potentially having either circular (fixed values for $e_i=0$ and $\omega_i=90^{\circ}$) or eccentric orbits. We also tested different types of trends, from the simplest linear trend (including a slope parameter) to a quadratic one (including a quadratic term) for the case of no planets (labelled as "0p*") and one-planet ("1p*") models. In total, 21 models with different levels of complexity are tested (from 0 to 3 planets). The labelling for the models follows the nomenclature Xp[P$_i$c][Z], where X is the total number of planets assumed in the system, P$_i$c are the identifiers of the planets with assumed circular orbits (e.g. "2p1c2c" if both planets "1" and "2" are assumed in circular orbits; or "2p1c" if only planet "1" is assumed in circular orbit), and the latter letter ("[Z]") indicates whether we are using a flat ("F") slope (i.e., assuming slope$ = 0$) or a quadratic ("Q") trend. 
 
\begin{figure*}
\centering
\includegraphics[width=0.48\textwidth]{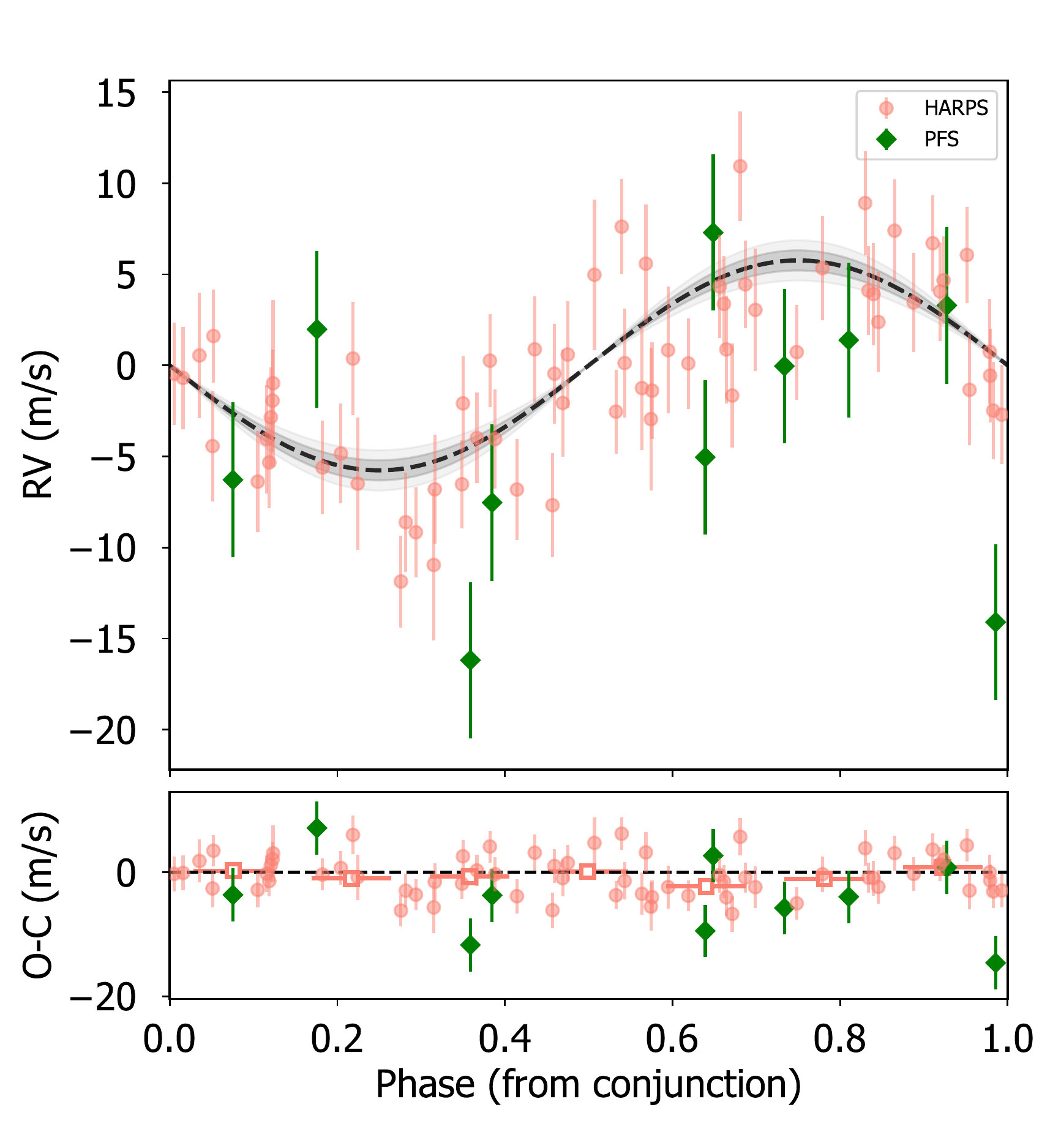}
\includegraphics[width=0.48\textwidth]{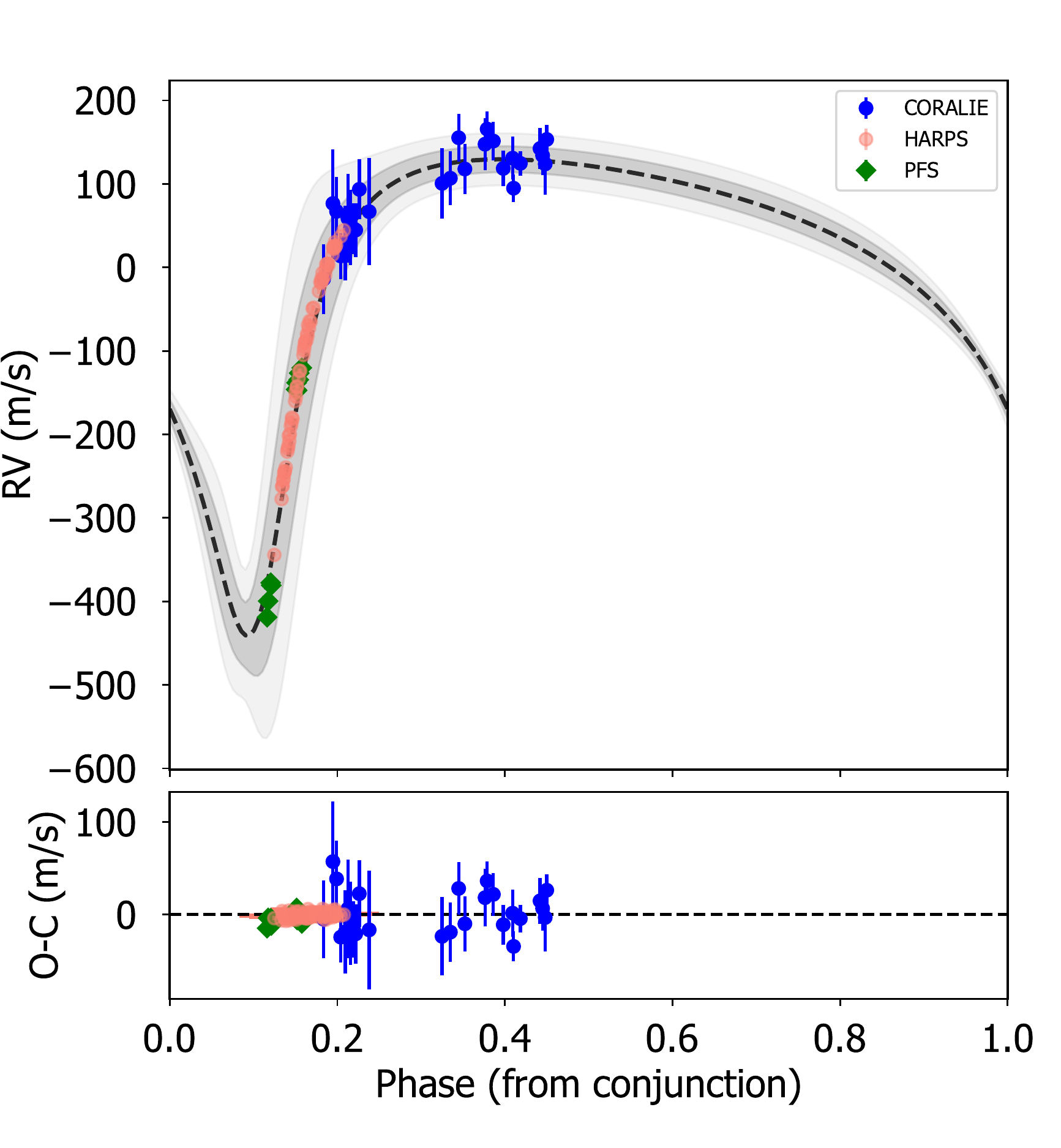}
\caption{Phase-folded radial velocity data from \toib{} (left) and \toic{} (right) after removing the signal from the other component in the system and the GP noise for the selected model (labelled "2p1cF") from the radial velocity analysis (see Sect.~\ref{sec:RVanalysis}). For \toib{} we just include HARPS and PFS data while removing CORALIE for clarity reasons given the larger uncertainties for that dataset. The gray shaded regions correspond to the region covered by models within 68.7\% (dark gray) and 95\% (light gray) of the confidence interval.}
\label{fig:RVphase_b}
\end{figure*}

 The Bayesian evidence is then used to select the simplest model that best represents our data. This is done by sorting the models by complexity and selecting the one with the highest evidence, assuming that a difference of 6 in logarithmic space (i.e., $\mathcal{B} = \Delta \ln{\mathcal{Z}}> 6$, \citealt{trotta08}) corresponds to a strong evidence in favor of the more complex model. Figure~\ref{fig:RVmodelcomp} shows this log-evidence for each of the 21 models. From this, we see that the model with a flat slope and two components in eccentric orbits ("2pF") has the largest evidence. However, the odds against the simpler model with the inner component in a circular orbit "2p1cF" is still not significant and we choose the latter as the preferred model informed by the current dataset. The evidence of this model against the no-planet model and the one-planet model is significantly larger ($\mathcal{Z}_{\rm 2p1cF}- \mathcal{Z}_{\rm 0p*} > +46$ and $\mathcal{Z}_{\rm 2p1cF}- \mathcal{Z}_{\rm 1p*} > +33$) thus supporting the existence of another Keplerian signal in the system other than the transiting planet. On the other side, the "2p1cF" model also has significantly larger evidence than the more complex three-planet models ($\mathcal{Z}_{\rm 2p1cF}- \mathcal{Z}_{\rm 3p*} > +31$). The phase-folded radial velocity curves of both components for the preferred model "2p1cF" are shown in Fig.~\ref{fig:RVphase_b} 
 The posterior distributions of all parameters involved in this analysis are presented in Table~\ref{tab:RValone}.
 
\subsection{Joint analysis}
\label{sec:joint}

We use the previous individual analysis of the light curve and radial velocity datasets as exploratory studies of the parameter space and for model comparison purposes. Based on the results obtained in these preliminary analysis, we now run a joint model including both the radial velocity and photometric light curves, from which we will obtain the final parameters of the system. In this case, the planetary and orbital parameters are parameterized to include the shared information between both techniques. That is, the Keplerians are modeled by a planet period, time of mid-transit, planet mass ($m_i$) and radius ($R_i$), orbital inclination ($i_i$), eccentricity and argument of the periastron. Besides this, as detailed in Sect.~\ref{sec:TESSanalysis} and Sect.~\ref{sec:RVanalysis}, we use the \texttt{RotationTerm} kernel with shared hyper-parameters (except for the amplitude) for the TESS light curves, radial velocity and BIS-span time series.

We proceed in a similar fashion as explained in Sect.~\ref{sec:RVanalysis} to explore the parameter space of this joint analysis. We reach convergence of the chains under this strategy and {confirm the planetary nature of the inner component and place the second component in the planet-to-brown-dwarf transition}. Unfortunately, the TESS observations do not cover the conjunction time of \toic{} and consequently we cannot conclude on its transiting nature. The final set of parameters (including the absolute mass for \toib{}) are presented in Table~\ref{tab:joint} and a summary of the main properties of the two  components is presented in Table~\ref{tab:JointSummary}.

\begin{table}
\setlength{\extrarowheight}{3pt}
\centering
\caption{Summary of the posterior distributions of the main physical and orbital properties of \toib{} and \toic{} from the joint analysis (Sect.~\ref{sec:joint}).}
\label{tab:JointSummary}
\begin{tabular}{ll}
\hline\hline
\multicolumn{2}{l}{\it \toib{}}  \\
\hline

Orbital period, $P_b$ [days]  & $1.8237305^{+0.0000020}_{-0.0000021}$ \\
Time of mid-transit, $T_{\rm 0,b}$ [days]  & $2459248.37709^{+0.00036}_{-0.00039}$ \\
Planet mass, $M_b$ [\Mearth{}]  & $9.1^{+1.0}_{-1.0}$ \\
Orbital inclination, $i_{\rm b}$ [deg.]  & $86.75^{+0.38}_{-0.41}$ \\
Planet radius, $R_b$ [\Rearth{}] & $2.765^{+0.088}_{-0.097}$ \\
Planet density, $\rho_{b}$ [$g\cdot cm^{-3}$]  & $2.34^{+0.39}_{-0.34}$ \\
Transit depth, $\Delta_{b}$ [ppt]  & $1.435^{+0.043}_{-0.043}$ \\
Orbit semi-major axis, $a_{b}$ [AU]  & $0.02636^{+0.00017}_{-0.00017}$ \\
Relative orbital separation, $a_{b}/R_{\star}$  & $8.47^{+0.21}_{-0.21}$ \\
Transit duration, $T_{\rm 14,b}$ [hours] & $1.519^{+0.018}_{-0.018}$ \\
Planet surface gravity, $g_{\rm b}$ [$m \cdot s^{-2}$] & $11.6^{+1.6}_{-1.5}$ \\
Impact parameter, $b_{b}/R_{\star}$ & $0.480^{+0.044}_{-0.051}$ \\
Incident Flux, $F_{\rm inc,b}$ [$F_{{\rm inc},\oplus}$]& $188^{+26}_{-24}$ \\
Equilibrium temperature, $T_{\rm eq,b}$ [K]  & $941^{+31}_{-31}$ \\

\hline
\multicolumn{2}{l}{\it \toic{}}  \\
\hline
Orbital period, $P_c$ [days]  & $1700^{+290}_{-280}$ \\
Time of mid-transit, $T_{\rm 0,c}$ [days] & $2460640^{+260}_{-260}$ \\
Minimum mass, $m_c\sin{i_c}$ [\Mjup{}]  & $11.3^{+1.1}_{-0.9}$ \\
Orbital eccentricity, $e_{\rm c}$  & $0.628^{+0.043}_{-0.036}$ \\
Arg. periastron, $\omega_{\rm c}$ [deg.] & $208.5^{+7.8}_{-7.3}$ \\
Orbit semi-major axis, $a_{c}$ [AU] & $2.52^{+0.27}_{-0.29}$ \\
Relative orbital separation, $a_{c}/R_{\star}$  & $806^{+91}_{-93}$ \\
Incident Flux, $F_{\rm inc,c}$ [$F_{{\rm inc},\oplus}$]  & $0.0208^{+0.0065}_{-0.0046}$ \\
Equilibrium temperature, $T_{\rm eq,c}$ [K] & $96.4^{+6.8}_{-5.8}$ \\

\hline\hline
\end{tabular}
\tablefoot{For a full description of all parameters involved and their priors, see Table~\ref{tab:joint}.}
\end{table}

\subsection{Transit timing variations for \toib{}}
\label{sec:TTVs}

\begin{figure}
\centering
\includegraphics[width=0.48\textwidth]{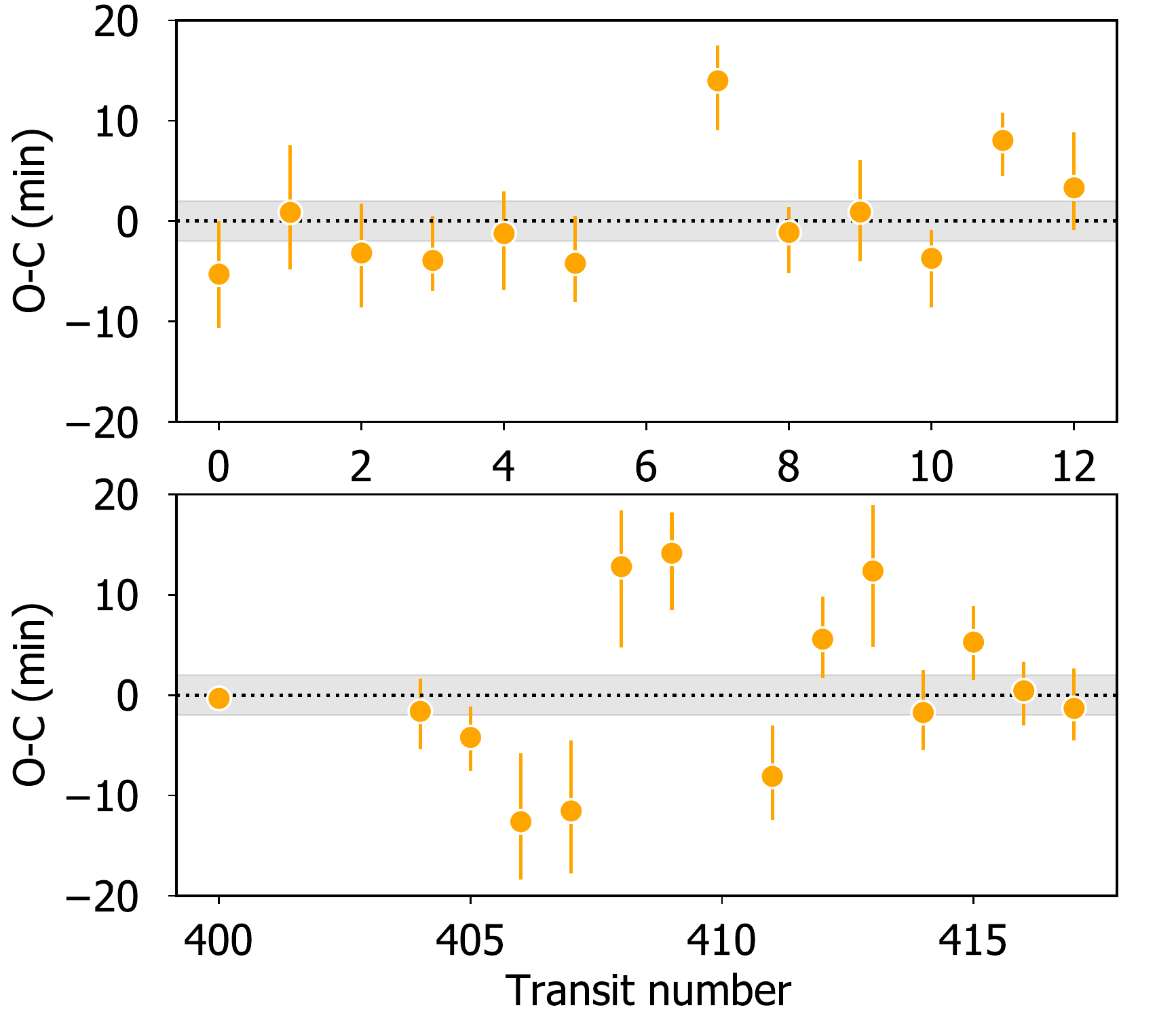}
\caption{Observed$-$Calculated diagram of the transit times of \toib{}. LCOGT transit corresponds to transit number \#400. The gray regions represent the 1$\sigma$ uncertainties of the ephemeris reported in Table~\ref{tab:joint}.}
\label{fig:TTVs}
\end{figure}

Given the relatively large number of transits of \toib{} in our photometric dataset, we can study the presence of additional planets in the system based on the gravitational pull on this planet. We estimate the central times of individual transit events of \toib{} based on deviations from a single-planet model. To that end, we use the results of the joint modeling of Sect.~\ref{sec:joint} (Table~\ref{tab:joint}) as priors in this analysis. As in the global analysis, we also used GPs to detrend the TESS light curves. On the contrary, we used the already-detrended LCOGT/MusCat light curves to estimate a single transit mid-time out of the simultaneous modeling of the four-filter light curves. We use the \texttt{juliet} package\footnote{\texttt{juliet} core tools used in this analysis are: \texttt{batman} \citep{batman}, \texttt{starry} \citep{starry}, \texttt{dynesty} \citep{dynesty}, and Gaussian Processes with \texttt{george} \citep{george} and \texttt{celerite} \citep{celerite}.} \citep{juliet} to perform this analysis. By doing this, we estimate the central times of the 25 TESS transits and 1 of the LCOGT light curves (Table~\ref{tab:transit_times}). We compare these observed times with the ephemeris reported in Table~\ref{tab:joint}. The resulting observed {\it minus} calculated (O$-$C) diagram of the transit times of TOI-969\,$b$ are shown in Fig.~\ref{fig:TTVs}. The RMS of the O-C points is 7.1\,min with a maximum deviation of 14.13\,min.  We noticed that the amplitude of the timing deviations are larger in the transits observed in the last TESS Sector (bottom panel in Fig.~\ref{fig:TTVs}). It is however unlikely that these deviations are related to the stellar activity as it can be seen in Fig.~\ref{fig:TESS}, the stellar activity is lower in this TESS Sector.

\begin{table}[]
    \centering
    \setlength{\extrarowheight}{3pt}
    \small
    \caption{Transit times of \toib{} estimated from the TESS and LCOGT light curves. }
    \label{tab:transit_times}
    \begin{tabular}{cccc}
    \hline
    Transit  & T$_c$ & O-C & Instrument\\
    number      & BJD-2\,450\,000 & [min] & \\
    \hline\hline
 0  & 8493.34879 $^{+0.0037}_{-0.0037}$ & -5.28  & TESS_S07 \\
 1 & 8495.17679 $^{+0.0046}_{-0.0040}$   & 0.87 & TESS_S07  \\
 2 & 8496.99771 $^{+0.0034}_{-0.0038}$   & -3.18 & TESS_S07  \\
 3 & 8498.82093 $^{+0.0031}_{-0.0021}$   & -3.92 & TESS_S07  \\
 4 & 8500.64652 $^{+0.0029}_{-0.0039}$ & -1.23  & TESS_S07 \\
 5 & 8502.46819 $^{+0.0033}_{-0.0027}$  & -4.20 & TESS_S07  \\
 6 & 8506.12828 $^{+0.0025}_{-0.0034}$ & 13.98  & TESS_S07 \\
 7 & 8507.94152 $^{+0.0017}_{-0.0028}$  & -1.12 & TESS_S07 \\
 8 & 8509.76667 $^{+0.0036}_{-0.0034}$ & 0.91  & TESS_S07 \\
 9 & 8511.58719 $^{+0.0020}_{-0.0034}$  & -3.70 & TESS_S07  \\
 10 & 8513.41908 $^{+0.0019}_{-0.0025}$ & 8.04  & TESS_S07 \\
 11 & 8515.23953 $^{+0.0038}_{-0.0030}$  & 3.32 & TESS_S07  \\
 12 & 9230.13869 $^{+0.0023}_{-0.0026}$ & -1.62 & TESS_S07  \\
 400 & 9222.84466 $^{+0.00042}_{-0.00040}$ & -0.34 & LCOGT \\
  404 & 9231.96062 $^{+0.0022}_{-0.0023}$  & -4.22 & TESS_S34  \\
  405 & 9233.77852 $^{+0.0048}_{-0.0040}$ & -12.62  & TESS_S34 \\
  406 & 9235.60299 $^{+0.0049}_{-0.0043}$  & -11.55  & TESS_S34 \\
  407 & 9237.44362 $^{+0.0039}_{-0.0056}$ & 12.79  & TESS_S34 \\
  408 & 9239.26829 $^{+0.0028}_{-0.0039}$  & 14.13 & TESS_S34  \\
  409 & 9242.90031 $^{+0.0035}_{-0.0030}$ & -8.10  & TESS_S34 \\
  411 & 9244.73353 $^{+0.0030}_{-0.0027}$ & 5.56  & TESS_S34 \\
  413 & 9246.56198 $^{+0.0046}_{-0.0053}$ & 12.35  & TESS_S34 \\
   414 & 9248.37592 $^{+0.0029}_{-0.0026}$ & -1.74 & TESS_S34  \\
   415 & 9250.20452 $^{+0.0025}_{-0.0026}$ & 5.27  & TESS_S34 \\
   416 & 9252.02489 $^{+0.0020}_{-0.0024}$ & 0.43  & TESS_S34 \\
   417 & 9253.84740 $^{+0.0028}_{-0.0022}$ & -1.33 & TESS_S34  \\
 
\hline\hline
\end{tabular}
\tablefoot{The values of the O$-$C column correspond to the points shown in Fig.~\ref{fig:TTVs}, which were obtained using the ephemeris reported in Table~\ref{tab:joint} and described in Sect.~\ref{sec:TTVs}.}
\end{table}

Based on the constraints of these transit timing results, we explore the limits that can be placed on the mass of a potential unseen companions to \toib{}. For this purpose, we used the \texttt{TTV2Fast2Furious} code\footnote{https://github.com/shadden/TTV2Fast2Furious} \citep{Hadden2019}. The resulting upper limits on the companion are not very constraining, ruling out planets with masses above $>$100~\Mearth{} for periods larger than twice the orbital period of \toib{}. Better constraints can be placed with an extended timing monitoring of additional \toib{} transits.

\section{Discussion}
\label{sec:Discussion}

\subsection{\toib{}: a new hot-Neptune in the desert}
\label{sec:hot-neptune}

\begin{figure*}
\centering
\includegraphics[width=1\textwidth]{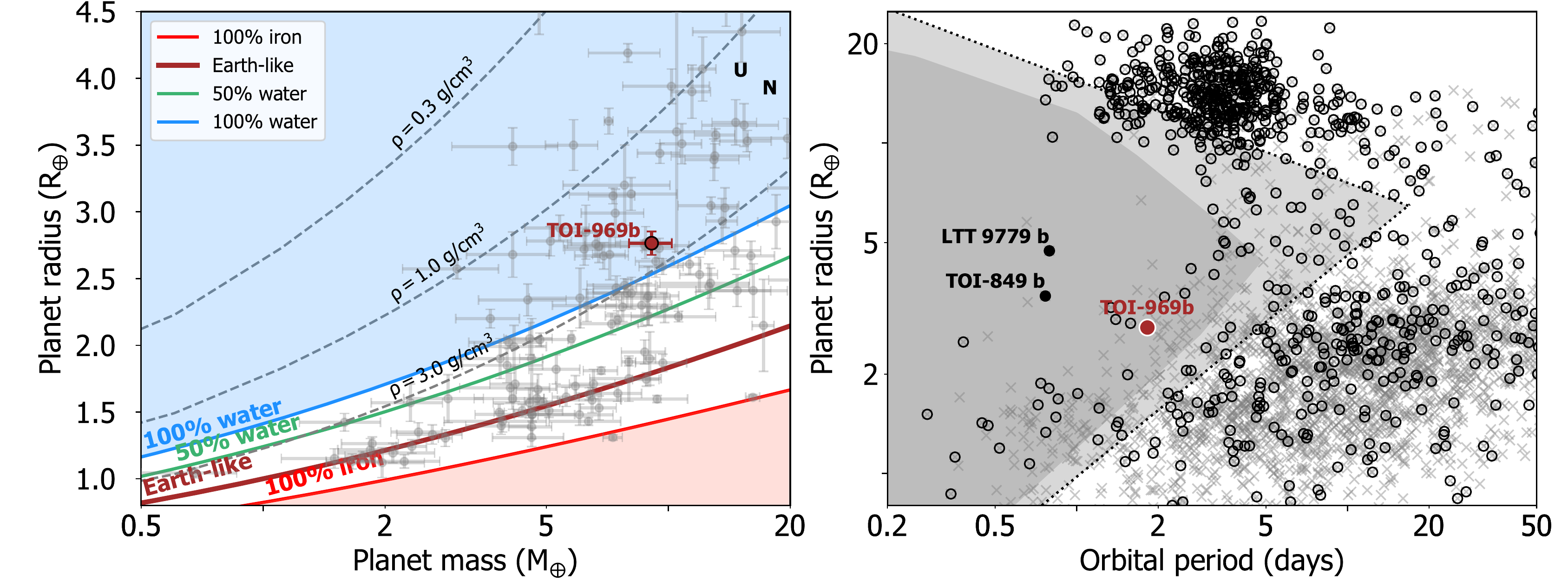}
\caption{Comparative properties of \toib{}. \textbf{Left}: Mass-radius diagram of known exoplanets with measured mass precisions better than 30\% (gray symbols) and the location of \toib{}. The bulk density lines corresponding to different compositions from \cite{zeng19} are shown as solid traces, and the dashed lines correspond to iso-densities of 0.3, 1, and 3 g$\cdot$cm$^{-3}$. \textbf{Right:} Period-radius diagram of known exoplanets (gray crosses) highlighting those with measured masses (open black circles). The light grey shaded region represents the mean boundaries of the hot-Neptune desert derived by \cite{mazeh16} and the dark grey shaded region corresponds to the interior envelope of the 1$\sigma$ boundaries. The location of \toib{} is marked in red, as well as the location of other two planets in the hot-Neptune desert, namely TOI-849\,b \citep{armstrong20} and LTT\,9779\,b \citep{jenkins20}.}
\label{fig:MR}
\end{figure*}

The derived properties of planet \toib{} showed in the previous sections place this world in the so-called mini-Neptune regime in terms of mass ($m_b=9.1^{+1.1}_{-1.0}$~\Mearth{}) and radius ($R_b=2.765^{+0.088}_{-0.097}$~\Rearth{}), see Fig.~\ref{fig:MR}. Its relatively short period also places this planet in the boundary of the hot-Neptune desert \citep{mazeh16} and possesses similar properties to the iconic GJ\,1214\,b \citep{charbonneau09}. 

{As for most Neptune-like planets, for a planet of the mass and radius of \toib{} and a H/He dominated atmosphere, the atmospheric loss by blowoff is energy limited \citep{owen16}. In this case, the atmospheric mass-loss rate is \citep{erkaev07,owen13}:}
~
\begin{equation}
\dot{M}=\epsilon \frac{\pi F_{\mathrm{XUV}} R_{\mathrm{p}}^3}{G M_{\mathrm{p}}},
\end{equation}
~
\noindent {where $F_{\mathrm{XUV}}$ is the XUV flux received by the planet, $G$ the gravitational constant, and $\epsilon$ is an efficiency parameter. We approximate the XUV luminosity by the analytical fit obtained by \cite{sanz-forcada11}, and we estimate $\epsilon\simeq0.07$ from \cite{owen12}. This yields a present-day mass-loss rate of 0.03 $\Mearth/$Gyr, assuming the age of the star is 2.03 Gyr. Following the approach of \cite{aguichine21}, the total mass lost by the planet can be estimated by integrating $\dot{m}$ over time assuming that only the XUV luminosity changes, and that the planet properties remained roughly constant during its evolution. In this case, we find that TOI-969 b would have lost $\sim 0.42\Mearth$ of H/He during its evolution. The possibility that TOI-969 b formed with a H/He envelope of mass greater than $\sim 0.42\Mearth$ ($\sim 4\%$ by mass) cannot be excluded. It is thus possible that TOI-969 b is still subject to atmospheric escape, and could be entirely stripped of its H/He envelope. However, the presence of an atmosphere made of heavier volatiles, such as He, O$_2$, H$_2$O and other (see \citealt{hu15,bolmont17,aguichine21,ito21}, respectively), provides a more natural explanation to how TOI-969 b retained its volatile envelope, since these molecules have smaller escape rates \citep{owen12,ito21}.}

\toib{} is hence another excellent target for testing theories of atmospheric evaporation \citep{lecavelier07,owen19}. Interestingly, its transmission spectroscopy metric (TSM, \citealt{kempton18}) corresponds to ${\rm TSM} = 93\pm18$, hence being one of the best targets in this regime for atmospheric purposes, especially with the James Webb Space Telescope (JWST). 

\subsection{Internal structure of \toib{}}
\label{sec:internal}

\begin{figure}
\centering
\includegraphics[width=0.5\textwidth{}]{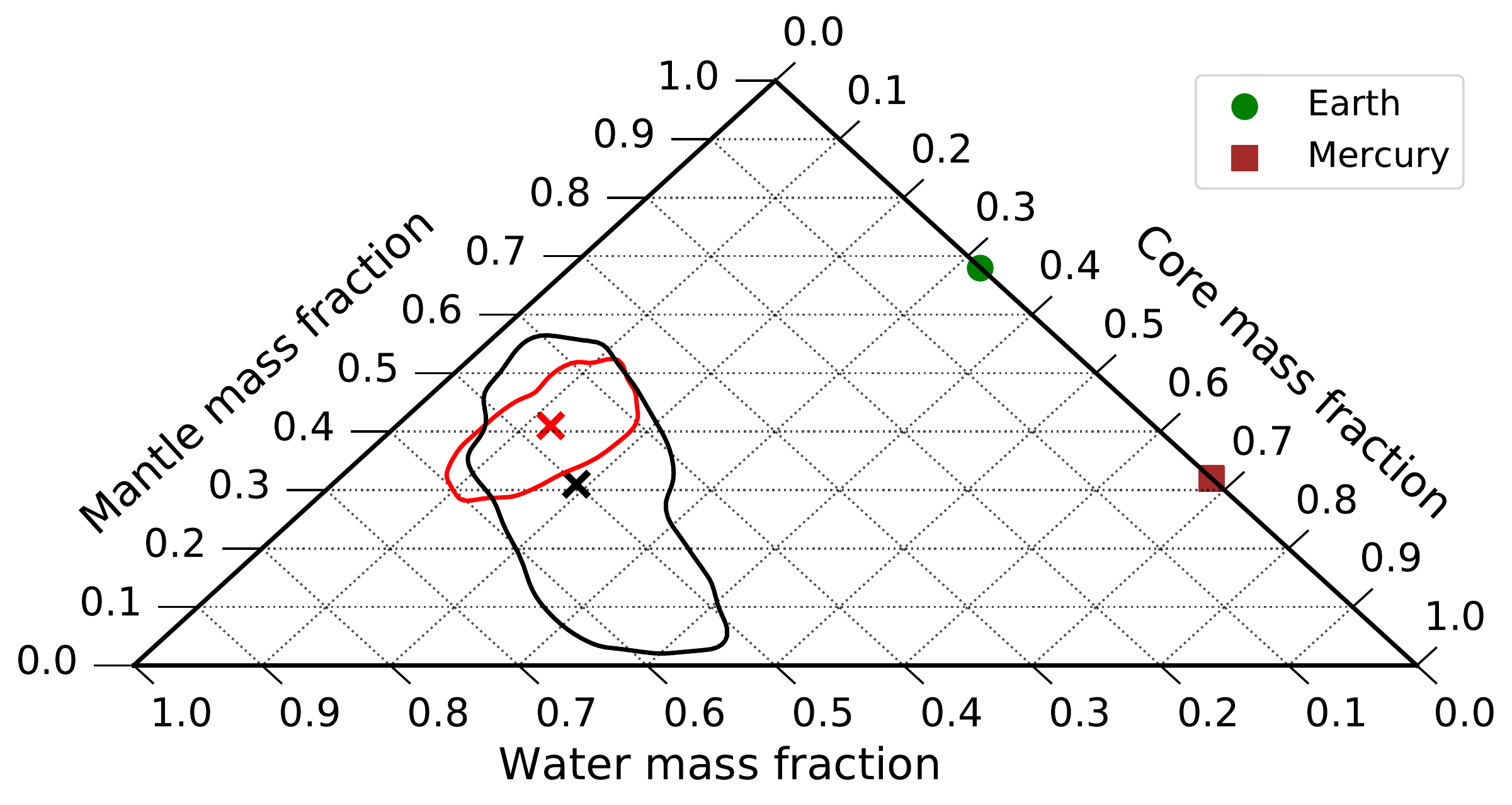}
\caption{Ternary diagram with the 1$\sigma$ confidence region of \toib{} in scenario 1 (black), where only the mass and the radius are considered as input data, and in scenario 2 (red), where the stellar abundances are also included as input data in the MCMC interior structure analysis. The mantle mass fraction (MMF) is defined as MMF = 1 - CMF - WMF. The green dot and red square indicate the position of Earth and Mercury in the ternary diagram, respectively.}
\label{fig:ternary_comp}
\end{figure}

{By using the precise mass and radius determinations of \toib, we can study its internal structure}. We performed a Markov chain Monte Carlo (MCMC) Bayesian analysis \citep{Dorn15} of its internal composition by using the interior structure model introduced in \cite{mousis20} and \cite{brugger17}, which comprises three layers: a Fe-rich core, a silicate rich mantle and a water layer.  With an irradiance temperature\footnote{{Irradiance temperature is the term used in atmospheric physics to name the equilibrium temperature at zero albedo}.} {of the order of $10^{3}$~K}, \toib{} can present vapor and supercritical\footnote{{Planets that are highly irradiated and present a volatile envelope reach high temperatures at the bottom of this layer, which corresponds to high pressures. Water is the second most abundant volatile after H/He. Its phase diagram (see, for example, Fig. 1 in \citealt{mousis20}) shows that for temperatures around $10^3$~K, water will reach the supercritical phase (decreasing its density compared to liquid or ice phases) when the pressure is approximately 300 bar.}} phases if water is found at its surface. Therefore we couple an atmosphere-interior model that calculates the surface conditions and the contribution of the atmosphere to the total radius, which in the case of low-density, warm planets is significant, as shown by \cite{acuna21}. We consider two scenarios to obtain the interior structure of \toib{}: scenario 1, where only the mass and radius of the planet (from Table~\ref{tab:joint}) are considered as inputs to the MCMC analysis; and scenario 2, where the planetary mass and radius, and the stellar Fe/Si mole ratios (see Table~\ref{tab:StePar}) are the input data\footnote{{\cite{adibekyan21} recently found a relation between the composition of low-mass planets and their host stars, but the relation was not one-to-one.}}. To compute the Fe/Si and Mg/Si mole ratios with the stellar abundances, we follow the approach described in \cite{brugger17} and \cite{sotin07}, and obtain Fe/Si = 0.77$\pm$0.23 and Mg/Si = 1.01$\pm$0.44. The output of the MCMC analysis are the posterior distributions functions (PDF) of the core mass fraction (CMF), the water mass fraction (WMF) and the atmospheric parameters, which are the temperature at 300 bar, the planetary albedo and the atmospheric thickness from transit pressure (20 mbar, see \citealt{mousis20,grimm18}) to 300 bar\footnote{{The critical point of water is located at 220 bar, this is the pressure at which water transitions from vapour to supercritical. 300 bar is close enough to the pressure of the critical point to prevent the atmospheric model from taking over pressures where the opacity is very high.}}. We assume a water-rich atmosphere. Table  \ref{tab:comp_output} (upper part) shows the 1D, 1$\sigma$ confidence intervals of the MCMC output parameters and Fig.~\ref{fig:ternary_comp} presents the ternary diagram and the 1$\sigma$ confidence region for the internal structure of \toib{}. Under the assumption of absence of an H/He layer, \toib{} is a volatile-rich planet that could have up to 60\% of its mass in a water-rich volatile layer in both scenarios. The CMF is compatible with an Earth-like value in scenario 1, whereas in scenario 2 the CMF is lower due to a lower Fe/Si mole ratio of the host star compared to the solar value (Fe/Si$_{\odot}$ = 0.96). 

{H/He and water are the most abundant volatiles. In interior modelling, when including a volatile layer, it is widely acknowledged to assume that the main component of the envelope is either hydrogen, which is representative of a primordial atmosphere, or water, whose density represents better that of a secondary atmosphere. The interior structure model described above presents the implementation of a water layer. However, as we have described in Sect.~\ref{sec:hot-neptune} the presence of H/He in the current atmosphere of \toib{} cannot be discarded based on atmospheric loss. Moreover, the water layer model described above yields a maximum WMF of almost 60\% within 1$\sigma$ (WMF~=50\%~$\pm$~9\%). This is less than the maximum WMF found in Solar System bodies (i.e comets), which is 70-80\% \citep{mckay19}. However, it is still a high WMF compared to other low-mass exoplanets. Hence, we also explore the possibility of a H/He atmosphere instead of a water-dominated volatile layer.}

\begin{table}[h]
\setlength{\extrarowheight}{4pt}
\centering
\caption{Median and 1$\sigma$ confidence intervals of the interior and atmosphere modeling. MCMC output parameters (see Sect.~\ref{sec:internal}).}
\begin{tabular}{lcc}
\hline\hline
Parameter & Scenario 1\tablefootmark{a} & Scenario 2\tablefootmark{b} \\ 
\hline
\textit{Water as volatile only} & & \\
\hline
Core mass fraction, CMF & 0.19$\pm$0.16 & 0.12$\pm$0.04 \\
Water mass fraction, WMF & 0.50$\pm$0.09 & 0.47$\pm$0.09 \\
Fe-to-Si mole ratio, Fe/Si & 0.55$^{+1.07}_{-0.55}$ & 0.71$\pm$0.26 \\
Temp. at 300 bar, T$_{300}$ [K] & \multicolumn{2}{c}{4500} \\
Thickness at 300 bar [km] & 1179 $\pm$ 109 & 972 $\pm$ 95 \\
Albedo, a$_{p}$& \multicolumn{2}{c}{0.20$\pm$0.01} \\
Core+Mantle radius, [R$_{p}$] & 0.48$\pm$0.07 & 0.52$\pm$0.05 \\ 
\hline
\textit{H/He atmosphere} & & \\
\hline
Core mass fraction, CMF & 0.70$\pm$0.30 & 0.28$^{+0.05}_{-0.09}$ \\
H/He mass fraction, $x_{H/He}$ [$\times 10^{-3}$] & 3.8$\pm$2.0 & 1.4$^{+0.2}_{-0.3}$ \\
Fe-to-Si mole ratio, Fe/Si & 17.0$^{+25.8}_{-17.0}$  & 0.76$^{+0.21}_{-0.26}$ \\
Thickness, z$_{atm}$ [km] & 8185 $^{+574}_{-1480}$  & 5844 $^{+908}_{-799}$ \\ 
\hline
\hline

\end{tabular}
\label{tab:comp_output}
\tablefoot{The results are shown for the two different compositional scenarios and for the two models considered (water as the only volatile - upper part- and a H/He atmosphere -bottom part).\\
\tablefoottext{a}{Only the mass and radius of the planet are considered as inputs.}\\
\tablefoottext{b}{Besides the planetary mass and radius, alse the stellar Fe/Si mole ratios (see Table~\ref{tab:StePar}) are included as input data.}}
\end{table}

\begin{figure}[]
\centering
\includegraphics[width=0.5\textwidth{}]{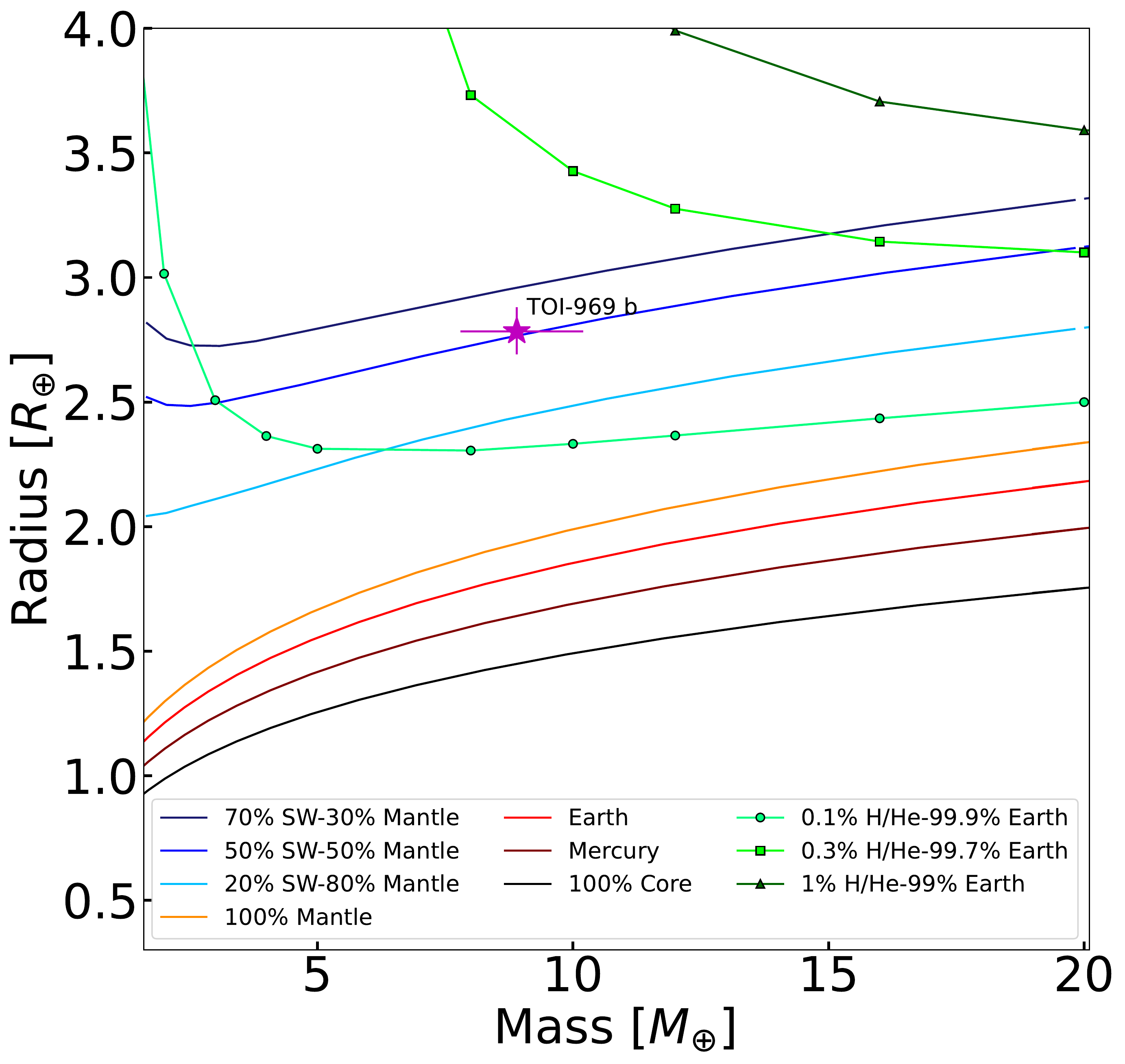}
\caption{Mass-radius relationships for supercritical water (SW) planets \citep{acuna2021,mousis20}, planets with Earth-like cores and H/He atmospheres \citep{zeng19} and dry planets with different Fe contents \citep{brugger17}. We assume high {chemical} equilibrium temperatures for supercritical and H/He volatile layers of 1200 K and 2000 K, respectively. The position of TOI-969 b is indicated as a magenta star.}
\label{fig:mrrel}
\end{figure}

To do this, we combine our interior structure model (only the Fe-rich core and the mantle) with the mass-radius relations of \cite{zeng19}. We obtain the atmospheric thickness by subtracting the mass-radius relationship of an Earth-like bulk with a H/He envelope minus that of a bare Earth-like core, both as provided by \cite{zeng19}\footnote{\url{https://lweb.cfa.harvard.edu/~lzeng/planetmodels.html}}. This allows us to express the atmospheric thickness as a function of the surface gravity, $g_{0} = M/R^{2}$, and the H/He mass fraction, $z_{\rm atm} = z_{\rm atm}(g_{0},$x$_{H/He})$. The boundary surface condition for our interior model are $T_{\rm surf}$ = 2000 K and $P_{\rm surf}$ = 1 bar, because \cite{zeng19} consider a maximum irradiation temperature of 2000 K and an isothermal profile in the atmosphere. Table \ref{tab:comp_output} (bottom part) shows the observables and the compositional parameters retrieved from this second analysis assuming a H/He atmosphere. We observe that the H/He volatile mass fraction is four orders of magnitude lower that the water mass fraction obtained in the previous analysis, which is expected since water at high pressures is significantly more dense than H and He. The 1$\sigma$ confidence intervals of the H/He volatile mass fractions are situated between 0.1\% and 0.3\%, as expected from the position of \toib{} in the mass-radius diagram in Fig. \ref{fig:mrrel}. To reproduce the density of \toib{}, the H/He case does not need a volatile layer as massive as that of the water case since a H/He atmosphere is more expanded than a secondary atmosphere for a similar atmospheric mass. This results in a larger portion of the total mass constituted by the Fe-rich core when we consider a H/He atmosphere, yielding higher CMFs in both scenarios. In scenario 2, we can observe that the CMF is compatible within uncertainties with the Earth value (CMF$_{\oplus}$ = 0.32).

Finally, we applied the stoichiometric model of \cite{santos17} to determine the iron-mass fraction (which can be translated into core-mass fraction) of the planet building blocks in the protoplanetary disks of \toi{}. This model is based on the chemical abundances of this star listed in Table~\ref{tab:StePar}, and assuming that C and O abundances for the two stars scale with metallicity. We find an iron mass fraction of the planet building blocks of $33.1 \pm 5.2$~\%. This value is compatible with that predicted by this model for the solar system planet building blocks (33.2\%, \citealt{santos15}).

\subsection{Prospects for atmospheric studies with JWST}
\label{sec_JWST}

We simulated synthetic spectra for a range of atmospheric scenarios and instrumental setups. We adopted Tau-REx~III \citep{al-refaie2021} to compute the model-atmosphere spectra, using the stellar parameters from Table~\ref{tab:StePar} (arithmetic average) and the planetary parameters from Table~\ref{tab:joint}. Our test bench models assume atmospheric chemical equilibrium (ACE, \citealp{agundez2012}) with isothermal profile at the equilibrium temperature, scaled (1$\times$ and 100$\times$) solar abundances, collisionally induced absorption (CIA) by H$_2$-H$_2$ and H$_2$-He \citep{abel2011,abel2012,fletcher2018}, and Rayleigh scattering. We show two cloud-free models with different metallicities, and a third model with solar metallicity and an optically thick cloud deck with top pressure of 100 Pa. The cloud-free spectra present molecular absorption features of 100-300 ppm, which are slightly smaller by a factor of $\sim$2 in the case with enhanced metallicity. The cloudy model also presents smaller absorption features due to the suppression of contributions from deeper atmospheric layers. A flat spectrum due to higher altitude (lower top pressure) clouds cannot be excluded for \toib{}.

\begin{figure*}
\centering
\includegraphics[width=\hsize]{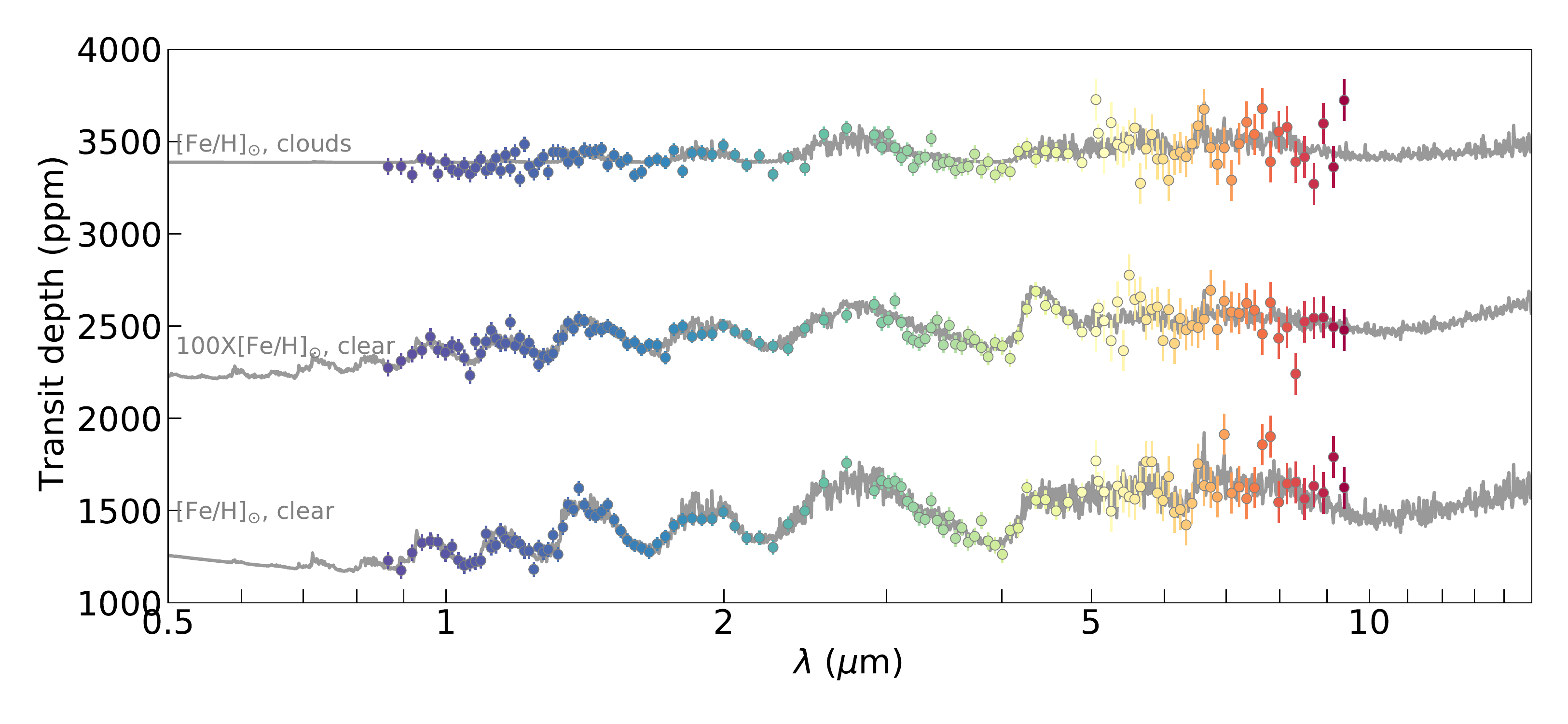}
\caption{Synthetic transmission spectra of \toib{}. Fiducial models assuming a cloud-free atmosphere with solar abundances (bottom), cloud-free atmosphere with enhanced metallicity by a factor of 100 (middle), and atmosphere with solar abundances and optically-thick cloud deck with top pressure of 100 Pa (top). Estimated uncertainties are shown for the observation of one transit with JWST NIRISS-SOSS, NIRSpec-G395M, and MIRI-LRS configurations. {Simulated data is color coded with wavelength.}}
\label{Fig:TOI969b_JWSTspectra}
\end{figure*}

We used \texttt{ExoTETHyS}\footnote{\url{https://github.com/ucl-exoplanets/ExoTETHyS}} \citep{morello2021} to compute bin-averaged spectra, taking into account the spectral response of the JWST instruments, noise scatter and error bars. We simulated JWST spectra for the NIRISS-SOSS (0.6--2.8 $\mu$m), NIRSpec-G395M (2.88--5.20 $\mu$m) and MIRI-LRS (5--12 $\mu$m) instrumental modes. The wavelength bins were specifically determined to have similar counts, leading to nearly uniform error bars per spectral point. In particular, we set a median resolving power of $R\sim 50$ for the NIRISS-SOSS and NIRSpec-G395M modes, and bin sizes of $\sim$0.1--0.2$\mu$m for the MIRI-LRS. The error bars have been calculated for a single visit of twice the transit duration in each instrumental mode, scaled by the inverse of the observing efficiency estimated with the Exoplanet Characterization Toolkit  (ExoCTK\footnote{\url{https://exoctk.stsci.edu}}, \citealp{bourque2021_exoctk}), and a factor 1.2 to account for correlated noise. We obtained error bars of 35--50\,ppm per spectral point for the  NIRISS-SOSS and NIRSpec-G395M modes, and 110--114 for the MIRI-LRS bins. These numbers suggest that a single transit observation is sufficient to sample the molecular absorption features in case of a clear atmosphere, even with 100$\times$ solar metallicity. A single NIRSpec-G395M observation could also be sufficient to detect the absorption features in a cloudy scenario, if the top pressure is higher than 100\,Pa (see Fig.~\ref{Fig:TOI969b_JWSTspectra}).

\subsection{\toic{}: A massive eccentric cold component in the planetary-to-brown-dwarf transition}

\begin{figure}
\centering
\includegraphics[width=0.5\textwidth{}]{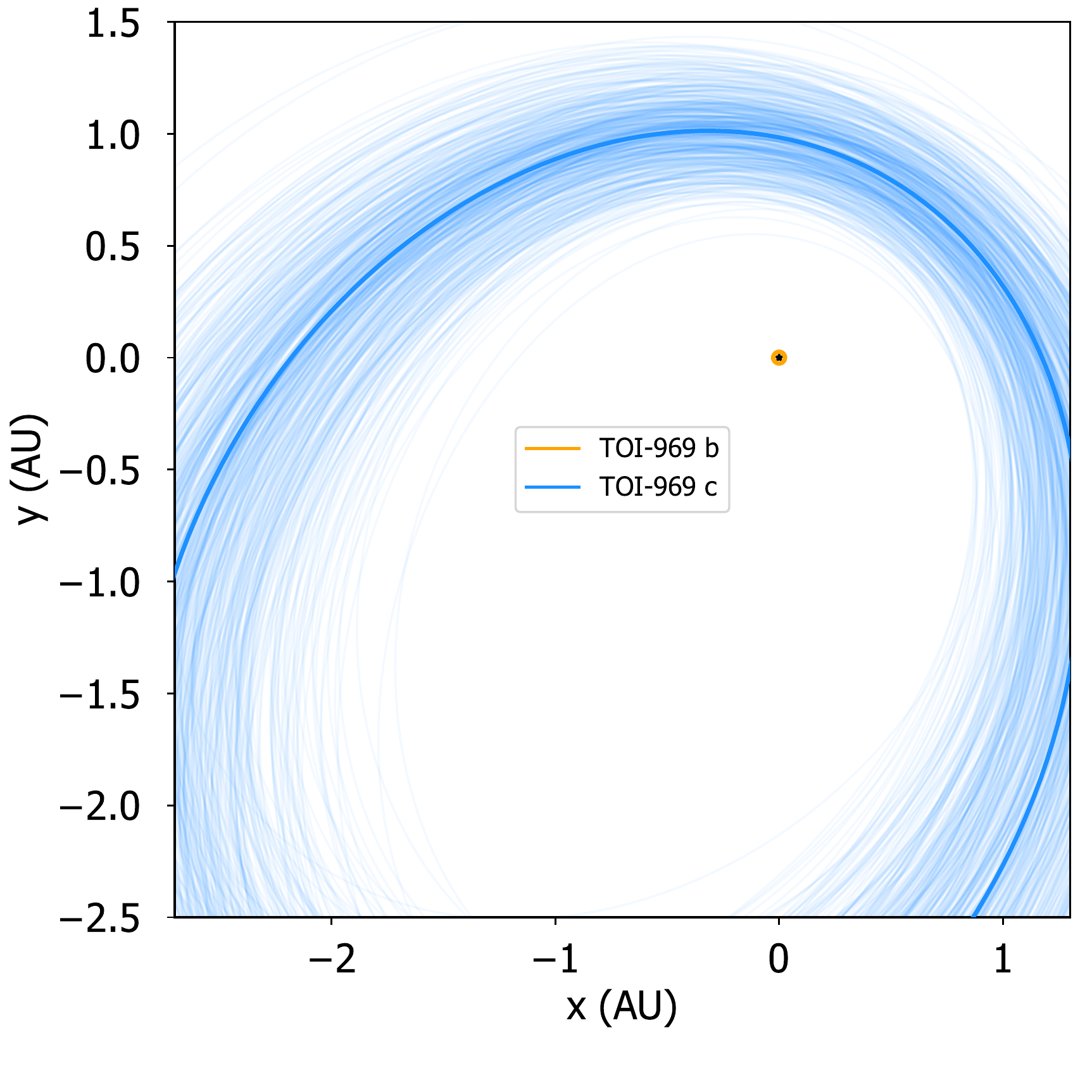}
\caption{Schematic view of the planetary system \toi{}, with the Earth direction towards the bottom of the panel. A total of 500 randomly selected orbits from the posterior distribution of the joint model (see Sect.~\ref{sec:joint}) are drawn for \toib{} (orange) and \toic{} (blue). The orbit corresponding to the median solution for each parameter is displayed as a solid thicker line of the corresponding color for each planet.}
\label{fig:orbit}
\end{figure}

 Our RV analysis strongly favors the presence of an outer companion at 2.4 AU in a high eccentric orbit that we call \toic{}. With a minimum mass of $11.3^{+1.1}_{-0.9}$~\Mjup{} and a derived eccentricity of $e_c=0.628^{+0.043}_{-0.036}$, our analysis places this planetary system as an ideal case for dynamical studies of planet migration and formation. Figure~\ref{fig:orbit} displays the orbital shape of the system. These high eccentricities are expected for massive planets as demonstrated in \cite{adibekyan13} and previously predicted by numerical simulations (e.g., \citealt{papaloizou01,bitsch13}). In particular, one of the main scenarios to explain the current population of eccentric cold giant planets is that of planet-disk interactions during the planet formation phase \citep{papaloizou01,kley06,bitsch13}. As discussed in \cite{bitsch20}, in the case of very massive gas giants with more than 5~\Mjup{}, the gaps opened by these massive bodies are so deep that they prevent the Lindblad resonances that allow the damping of high eccentricities. This is one possibility to explain the large eccentricity of \toic{} in the absence of additional massive bodies in the system. The alternative scenario consists of planet-planet scattering. This has been tested by different authors using different approaches and configurations of the forming planetary system (e.g., \citealt{juric08,raymond09a,bitsch20}) reaching to the conclusion that planetary systems including a massive eccentric giant should not harbor any inner super-Earths. By contrast, \cite{zhu18} and \cite{bryan19} found that there is a positive correlation between the presence of hot super-Earths and cold Jupiters in the same system. This is then still an open question. 

The presence of the hot mini-Neptune \toib{} at such close-in orbit is thus intriguing in this regard, as it seems to be an exception to the planet-planet scattering scenario, which should have destabilised the inner planet and removed it from the system. The alternative would then be that \toib{} formed inner to the cold giant and subsequently migrated towards its current location through type-I migration. However, the expected location of the snow line for the proto-planetary disk of this late-K dwarf star is at around 1 AU, too close to the current location of \toic{} to have had enough dynamical room to form a Neptune-like planet as \toib{}. Consequently, the formation of this system and its evolution until reaching its present configuration remains open. Thus, a complete re-arrangement of the initial planetary configuration cannot be discarded.

\section{Conclusions}
\label{sec:Conclusions}

In this paper we have confirmed the planetary nature of \toib{}, a planet candidate detected by the TESS mission through the transit technique. Additional transits of \toib{} were also observed from the ground using LCOGT and MEarth telescopes. We used high-spatial resolution images to discard possible contaminant sources within the TESS aperture as well as an intense RV monitoring with different instruments (namely HARPS, PFS and CORALIE). These RV time series revealed the presence of a long-term signal compatible with a new planet in the system.

We analyse this large dataset and conclude that \toib{} is a hot ($P_b\sim 1.82$~days) mini-Neptune with a radius of $2.765^{+0.088}_{-0.097}$~\Rearth{} and a mass of $9.1^{+1.1}_{-1.0}$~\Mearth{}. These properties place this planet within the lower boundary of the hot-Neptune desert. The internal structure analysis also suggests a core mass fraction of $12\pm 4$~\% and a large fraction of its mass contained in a potential atmospheric layer. The simulations we performed suggest this atmosphere could be characterized with JWST in a single visit, thus putting this target in the priority list for sub-Neptune planets. The transit times of this planet also display some deviations from a strictly periodic signal at the 10-min level. However, the precision from the TESS data is not sufficient to properly assess a TTV analysis and so infer the possible presence of additional planets, at least not better than with the RV data. 

The long timespan radial velocity time series indicate the presence of a second component in the system (\toic{}) in a long-period orbit with $P_c=1700^{+290}_{-280}$~days. Our two-planet model is here favored against a simpler one-planet model plus a quadratic trend. This second body has a minimum mass of $11.3^{+1.1}_{-0.9}$~\Mjup{} (in the transition between the planetary and brown dwarf domain) and a large eccentricity of $e_c=0.628^{+0.043}_{-0.036}$, constituting a relevant piece of the system in terms of dynamical stability and placing it as a key actor of \toi{} in moulding the current architecture of the planetary system. \toi{} thus becomes a test bench for planet migration theories and, in particular, for planet-planet scattering studies.

\begin{acknowledgements}
We thank the anonymous referee for their though revision of this manuscript that has improved its final quality.
J.L-B. acknowledges financial support received from "la Caixa" Foundation (ID 100010434) and from the European Unions Horizon 2020 research and innovation programme under the Marie Slodowska-Curie grant agreement No 847648, with fellowship code LCF/BQ/PI20/11760023. This research has also been partly funded by the Spanish State Research Agency (AEI) Projects No.PID2019-107061GB-C61 and No. MDM-2017-0737 Unidad de Excelencia "Mar\'ia de Maeztu"- Centro de Astrobiolog\'ia (INTA-CSIC).
R.L. acknowledges financial support from the Spanish Ministerio de Ciencia e Innovaci\'on, through project PID2019-109522GB-C52, and the Centre of Excellence "Severo Ochoa" award to the Instituto de Astrof\'isica de Andaluc\'ia (SEV-2017-0709).
D.J.A. acknowledges support from the STFC via an Ernest Rutherford Fellowship (ST/R00384X/1).
S.G.S acknowledges the support from FCT through Estimulo FCT contract nr.CEECIND/00826/2018 and POPH/FSE (EC).
G.M. has received funding from the European Union's Horizon 2020 research and innovation programme under the Marie Sk\l{}odowska-Curie grant agreement No. 895525.
S.H. acknowledges CNES funding through the grant 837319.
The French group acknowledges financial support from the French Programme National de Plan\'etologie (PNP, INSU).
This work is partly financed by the Spanish Ministry of Economics and Competitiveness through grants PGC2018-098153-B-C31.
We acknowledge the support by FCT - Funda\c{c}\~ao para a Ci\^encia e a Tecnologia through national funds and by FEDER through COMPETE2020 - Programa Operacional Competitividade e Internacionaliza\c{c}\~ao by these grants: UID/FIS/04434/2019; UIDB/04434/2020; UIDP/04434/2020; PTDC/FIS-AST/32113/2017 \& POCI-01-0145-FEDER-032113; PTDC/FISAST/28953/2017 \& POCI-01-0145-FEDER-028953.
P.J.W. is supported by an STFC consolidated grant (ST/T000406/1).
F.H. is funded by an STFC studentship.
T.A.S acknowledges support from the Funda\c{c}\~ao para a Ci\^encia e a Tecnologia (FCT) through the Fellowship PD/BD/150416/2019 and POCH/FSE (EC).
C.M.P. acknowledges support from the SNSA (dnr 65/19P).
This work has been carried out within the framework of the National Centre of Competence in Research (NCCR) PlanetS supported by the Swiss National Science Foundation. This project has received funding from the European Research Council (ERC) under the European Union's Horizon 2020 research and innovation programme (grant agreement SCORE No 851555).
O.D.S.D. is supported in the form of work contract (DL 57/2016/CP1364/CT0004) funded by national funds through Funda\c{c}\~ao para a Ci\^encia e a Tecnologia (FCT).
M.E. acknowledges the support of the DFG priority programSPP 1992 "Exploring the Diversity of Extrasolar Planets" (HA 3279/12-1).
A.O. is funded by an STFC studentship.
J.K. gratefully acknowledge the support of the Swedish National Space Agency (SNSA; DNR 2020-00104).
%
This work makes use of observations from the LCOGT network. This paper is based on observations made with the MuSCAT3 instrument, developed by the Astrobiology Center and under financial supports by JSPS KAKENHI (JP18H05439) and JST PRESTO (JPMJPR1775), at Faulkes Telescope North on Maui, HI, operated by the Las Cumbres Observatory.
Some of the observations in the paper made use of the High-Resolution Imaging instrument Zorro obtained under Gemini LLP Proposal Number: GN/S-2021A-LP-105. Zorro was funded by the NASA Exoplanet Exploration Program and built at the NASA Ames Research Center by Steve B. Howell, Nic Scott, Elliott P. Horch, and Emmett Quigley. Zorro was mounted on the Gemini North (and/or South) telescope of the international Gemini Observatory, a program of NSF's OIR Lab, which is managed by the Association of Universities for Research in Astronomy (AURA) under a cooperative agreement with the National Science Foundation. on behalf of the Gemini partnership: the National Science Foundation (United States), National Research Council (Canada), Agencia Nacional de Investigaci\'on y Desarrollo (Chile), Ministerio de Ciencia, Tecnolog\'ia e Innovaci\'on (Argentina),  Minist\'erio da Ci\^{e}ncia, Tecnologia, Inova\c{c}\~oes e Comunica\c{c}\~oes (Brazil), and Korea Astronomy and Space Science Institute (Republic of Korea).
We acknowledge the use of public TESS data from pipelines at the TESS Science Office and at the TESS Science Processing Operations Center.
Resources supporting this work were provided by the NASA High-End Computing (HEC) Program through the NASA Advanced Supercomputing (NAS) Division at Ames Research Center for the production of the SPOC data products.
The MEarth Team gratefully acknowledges funding from the David and Lucile Packard Fellowship for Science and Engineering (awarded to D.C.). This material is based upon work supported by the National Science Foundation under grants AST-0807690, AST-1109468, AST-1004488 (Alan T. Waterman Award), and AST-1616624, and upon work supported by the National Aeronautics and Space Administration under Grant No. 80NSSC18K0476 issued through the XRP Program. This work is made possible by a grant from the John Templeton Foundation. The opinions expressed in this publication are those of the authors and do not necessarily reflect the views of the John Templeton Foundation.
This research made use of Astropy, (a community-developed core Python package for Astronomy, \citealt{astropy:2013,astropy:2018}), SciPy \citep{scipy}, matplotlib (a Python library for publication quality graphics \citealt{matplotlib}), and numpy \citep{numpy}.
This research has made use of NASA's Astrophysics Data System Bibliographic Services. 
This research has made use of the SIMBAD database, operated at CDS, Strasbourg, France.
The authors wish to recognize and acknowledge the very significant cultural role and reverence that the summit of Mauna Kea has always had within the indigenous Hawaiian community.  We are most fortunate to have the opportunity to conduct observations from this mountain.

\end{acknowledgements}

%
%

\bibliographystyle{aa} 
\bibliography{../../../biblio2} 

\appendix
\section{ Additional tables}

\begin{table*}
\setlength{\extrarowheight}{3pt}
\centering
\caption{TESS light curves of TOI-969 (see Sect.~\ref{sec:tess}).}
\label{tab:TESSdata}
\begin{tabular}{cccccc}
\hline\hline
BJD & Flux (PDCSAP) & $\sigma_{\rm flux}$ & Flux det. (norm) & $\sigma_{\rm flux det.}$ & Sector \\ \hline
2458491.6829294283 & 1.001832 & 0.000371 & 0.999953 & 0.000371 & TESS07 \\
2458491.7037631846 & 1.002146 & 0.000369 & 1.000287 & 0.000369 & TESS07 \\
2458491.72459694 & 1.00138 & 0.000367 & 0.999544 & 0.000367 & TESS07 \\
2458491.745430695 & 1.002189 & 0.000365 & 1.000375 & 0.000365 & TESS07 \\
2458491.766264449 & 1.001059 & 0.000364 & 0.99927 & 0.000364 & TESS07 \\
2458491.7870982024 & 1.001779 & 0.000363 & 1.000014 & 0.000363 & TESS07 \\
2458491.8079319545 & 1.001815 & 0.000361 & 1.000076 & 0.000361 & TESS07 \\
2458491.828765706 & 1.001416 & 0.00036 & 0.999704 & 0.00036 & TESS07 \\
2458491.849599457 & 1.002108 & 0.000358 & 1.000425 & 0.000358 & TESS07 \\
2458491.870433206 & 1.001584 & 0.000357 & 0.999932 & 0.000357 & TESS07 \\
... & & & & &\\
\hline\hline
\end{tabular}
\tablefoot{The detrended light curve in the fourth column corresponds to the detrending applied by using the parameters from the joint analysis in Sect.~\ref{sec:joint}. Only the first 10 values are shown. The complete table is available through CDS (see link in the main title caption).}
\end{table*}

\begin{table*}
\setlength{\extrarowheight}{3pt}
\centering
\caption{Ground based photometric time series of TOI-969 from the LCOGT and MEarth facilities (see Sects.~\ref{sec:lco} and \ref{sec:MEarth}).}
\label{tab:GBdata}
\begin{tabular}{ccccccc}
\hline\hline
BJD & Flux (PDCSAP) & $\sigma_{\rm flux}$ & Flux det. (norm) & $\sigma_{\rm flux det.}$ & Airmass & Instrument \\ \hline
2459222.785142 & 1.005212 & 0.000751 & 1.000958 & 0.000751 & 2.293 & LCOHAL_gp \\
2459222.785448 & 1.002452 & 0.000748 & 0.99822 & 0.000748 & 2.285 & LCOHAL_gp \\
2459222.785755 & 1.003435 & 0.000751 & 0.999246 & 0.000751 & 2.277 & LCOHAL_gp \\
2459222.7861 & 1.00383 & 0.000745 & 0.999685 & 0.000745 & 2.267 & LCOHAL_gp \\
2459222.786406 & 1.002546 & 0.000739 & 0.99843 & 0.000739 & 2.259 & LCOHAL_gp \\
2459222.786714 & 1.004676 & 0.000736 & 1.000609 & 0.000736 & 2.251 & LCOHAL_gp \\
2459222.787021 & 1.002964 & 0.000733 & 0.998924 & 0.000733 & 2.243 & LCOHAL_gp \\
2459222.787373 & 1.002637 & 0.000736 & 0.998636 & 0.000736 & 2.234 & LCOHAL_gp \\
2459222.787683 & 1.004611 & 0.000736 & 1.000657 & 0.000736 & 2.226 & LCOHAL_gp \\
2459222.78799 & 1.004979 & 0.000748 & 1.001062 & 0.000748 & 2.218 & LCOHAL_gp \\
... & & & & &\\
\hline\hline
\end{tabular}
\tablefoot{The detrended light curve in the fourth column corresponds to the linear detrending with airmass applied by using the parameters from the joint analysis in Sect.~\ref{sec:joint}. Only the first 10 values are shown. The complete table is available through CDS (see link in the main title caption).}
\end{table*}

\begin{table*}
\setlength{\extrarowheight}{3pt}
\centering
\caption{Photometric time series from WASP (see Sect.~\ref{sec:wasp}).}
\label{tab:WASPdata}
\begin{tabular}{cccccc}
\hline\hline
BJD & Flux & $\sigma_{\rm flux}$ \\
\hline
2455155.610718 & 0.97743 & 0.00976 \\
2455155.611157 & 0.98893 & 0.00993 \\
2455155.617477 & 0.98432 & 0.01116 \\
2455155.617905 & 0.9859 & 0.01024 \\
2455155.624259 & 0.98289 & 0.00923 \\
2455155.624699 & 0.98068 & 0.00956 \\
2455155.630949 & 0.98545 & 0.01378 \\
2455155.631389 & 0.98309 & 0.01471 \\
2455155.63662 & 0.9821 & 0.01492 \\
2455155.637083 & 0.96412 & 0.01586 \\
... & & \\
\hline\hline
\end{tabular}
\tablefoot{Only the first 10 values are shown. The complete table is available through CDS (see link in the main title caption).}
\end{table*}

\begin{table*}
\small
\setlength{\extrarowheight}{3pt}
\centering
\caption{Radial velocity and activity indicators for all instruments.}
\label{tab:RVdata}
\begin{tabular}{lccccccc}
\hline\hline
JD & RV & BIS-SPAN & FWHM & Contrast & S-index & $\log{R^{\prime}_HK}$ & Inst.\\
(days) & (\kms) & (\kms) & (\kms) & (\%) & - & dex & \\
\hline
2459163.79166882 & $-5.8432 \pm 0.0018$ & $0.0480 \pm 0.0037$ & $6.6641 \pm 0.0037$ & $47.5 \pm 2.4$ & $1.153 \pm 0.021$ & $-4.3718 \pm 0.0081$ & HARPS \\
2459182.75297637 & $-5.7772 \pm 0.0022$ & $0.0644 \pm 0.0045$ & $6.6225 \pm 0.0045$ & $47.4 \pm 2.4$ & $1.021 \pm 0.027$ & $-4.426 \pm 0.012$ & HARPS \\
2459183.71415593 & $-5.7739 \pm 0.0025$ & $0.0683 \pm 0.0050$ & $6.6264 \pm 0.0050$ & $47.0 \pm 2.3$ & $1.032 \pm 0.028$ & $-4.421 \pm 0.012$ & HARPS \\
2459183.80873437 & $-5.7727 \pm 0.0030$ & $0.0640 \pm 0.0061$ & $6.6133 \pm 0.0061$ & $47.1 \pm 2.4$ & $0.993 \pm 0.037$ & $-4.437 \pm 0.017$ & HARPS \\
2459184.71221205 & $-5.7813 \pm 0.0027$ & $0.0510 \pm 0.0046$ & $6.6229 \pm 0.0054$ & $47.0 \pm 2.4$ & $1.048 \pm 0.031$ & $-4.414 \pm 0.013$ & HARPS \\
2459185.77198425 & $-5.7578 \pm 0.0023$ & $0.0531 \pm 0.0040$ & $6.6249 \pm 0.0046$ & $47.3 \pm 2.4$ & $1.077 \pm 0.027$ & $-4.402 \pm 0.011$ & HARPS \\
2459191.754039 & $-5.7141 \pm 0.0020$ & $0.0592 \pm 0.0046$ & $6.6709 \pm 0.0040$ & $47.5 \pm 2.4$ & $1.214 \pm 0.026$ & $-4.3494 \pm 0.0094$ & HARPS \\
2459192.67584257 & $-5.7268 \pm 0.0023$ & $0.0592 \pm 0.0050$ & $6.6572 \pm 0.0046$ & $47.6 \pm 2.4$ & $1.240 \pm 0.029$ & $-4.340 \pm 0.010$ & HARPS \\
2459193.75970319 & $-5.7209 \pm 0.0025$ & $0.0664 \pm 0.0037$ & $6.6772 \pm 0.0050$ & $47.6 \pm 2.4$ & $1.287 \pm 0.034$ & $-4.324 \pm 0.011$ & HARPS \\
2459195.70681719 & $-5.7156 \pm 0.0019$ & $0.0669 \pm 0.0040$ & $6.6480 \pm 0.0037$ & $47.7 \pm 2.4$ & $1.207 \pm 0.022$ & $-4.3517 \pm 0.0081$ & HARPS \\
... &... &... &... &... &... &... &... \\

\hline\hline
\end{tabular}
\tablefoot{Only the first 10 values are shown. The complete table is available through CDS (see link in the main title caption).}
\end{table*}

\begin{table*}
\setlength{\extrarowheight}{2.5pt}
\centering
\caption{Inferred and derived parameters from the light curve analysis (Sect.~\ref{sec:TESSanalysis}) for the one-planet model.}
\label{tab:LCalone}
\begin{tabular}{lll}
\hline\hline
Parameter & Priors & Posteriors \\
\hline
\textit{Planet modeled parameters} & &  \\
\hline

%
Orbital period, $P_b$ [days] & $\mathcal{G}$(1.8237371,0.0001) & $1.8237306\pm 0.0000021$ \\
Time of mid-transit, $T_{\rm 0,b}-2400000$ [days] & $\mathcal{G}$(59248.377773,0.1) & $59248.37713^{+0.00042}_{-0.00038}$ \\
Orbital inclination, $i_{\rm b}$ [deg.] & $\mathcal{U}$(40.0,90.0) & $86.71^{+0.37}_{-0.36}$ \\
Planet radius, $R_b$ [$R_{\oplus}$] & $\mathcal{U}$(0.0,20.0) & $2.773^{+0.089}_{-0.086}$ \\
Transit depth, $\Delta_{b}$ [ppt] & (derived) & $1.443^{+0.042}_{-0.042}$ \\
Orbit semi-major axis, $a_{b}$ [AU] & (derived) & $0.02636^{+0.00017}_{-0.00017}$ \\
Relative orbital separation, $a_{b}/R_{\star}$ & (derived) & $8.47^{+0.20}_{-0.20}$ \\
Transit duration, $T_{\rm 14,b}$ [hours] & (derived) & $1.515^{+0.018}_{-0.018}$ \\
Impact parameter, $b_{b}/R_{\star}$ & (derived) & $0.486^{+0.041}_{-0.046}$ \\
Incident Flux, $F_{\rm inc,b}$ [$F_{{\rm inc},\oplus}$] & (derived) & $529^{+26}_{-24}$ \\

\hline
\textit{Stellar parameters} & &  \\
\hline

Stellar radius, $R_{\star}$ [$R_{\odot}$] & $\mathcal{T}$(0.671,0.015,0.1,1.2) & $0.669^{+0.016}_{-0.015}$ \\
Stellar mass, $M_{\star}$ [$M_{\odot}$] & $\mathcal{T}$(0.734,0.014,0.1,1.2) & $0.734^{+0.014}_{-0.014}$ \\
Stellar luminosity, $L_{\star}$ [$L_{\odot}$] & (derived) & $0.367^{+0.017}_{-0.016}$ \\
Limb-darkening $u_1$, LCOHAL_gp & $\mathcal{T}$(0.592,0.05,0,1) & $0.595^{+0.026}_{-0.026}$ \\
Limb-darkening $u_2$, LCOHAL_gp & $\mathcal{T}$(0.141,0.05,0,1) & $0.141^{+0.025}_{-0.025}$ \\
Limb-darkening $u_1$, LCOHAL_ip & $\mathcal{T}$(0.539,0.05,0,1) & $0.538^{+0.025}_{-0.026}$ \\
Limb-darkening $u_2$, LCOHAL_ip & $\mathcal{T}$(0.155,0.05,0,1) & $0.154^{+0.025}_{-0.025}$ \\
Limb-darkening $u_1$, LCOHAL_rp & $\mathcal{T}$(0.7,0.05,0,1) & $0.706^{+0.025}_{-0.025}$ \\
Limb-darkening $u_2$, LCOHAL_rp & $\mathcal{T}$(0.078,0.05,0,1) & $0.083^{+0.026}_{-0.026}$ \\
Limb-darkening $u_1$, LCOHAL_zs & $\mathcal{T}$(0.416,0.05,0,1) & $0.414^{+0.026}_{-0.026}$ \\
Limb-darkening $u_2$, LCOHAL_zs & $\mathcal{T}$(0.195,0.05,0,1) & $0.193^{+0.026}_{-0.026}$ \\
Limb-darkening $u_1$, MEarth & $\mathcal{T}$(0.416,0.1,0,1) & $0.465^{+0.094}_{-0.095}$ \\

\hline
\textit{Instrument-dependent parameters} & &  \\
\hline
LC level TESS07 & $\mathcal{U}$(-20,20) & $-0.54^{+0.64}_{-0.76}$ \\
LC level TESS34 & $\mathcal{U}$(-20,20) & $-0.1^{+0.62}_{-0.70}$ \\
LC level LCOHAL_gp & $\mathcal{U}$(-20,20) & $5.69^{+0.18}_{-0.17}$ \\
LC level LCOHAL_ip & $\mathcal{U}$(-20,20) & $-0.48^{+0.19}_{-0.20}$ \\
LC level LCOHAL_rp & $\mathcal{U}$(-20,20) & $0.62^{+0.19}_{-0.19}$ \\
LC level LCOHAL_zs & $\mathcal{U}$(-20,20) & $-1.27^{+0.21}_{-0.20}$ \\
LC level MEarth & $\mathcal{U}$(-200,200) & $5.41^{+0.18}_{-0.17}$ \\
LC jitter TESS07 [ppm] & $\mathcal{U}$(0,4000) & $90^{+33}_{-28}$ \\
LC jitter TESS34 [ppm] & $\mathcal{U}$(0,4000) & $93^{+28}_{-56}$ \\
LC jitter LCOHAL_gp [ppm] & $\mathcal{U}$(0,4000) & $882^{+61}_{-44}$ \\
LC jitter LCOHAL_ip [ppm] & $\mathcal{U}$(0,4000) & $1588^{+36}_{-48}$ \\
LC jitter LCOHAL_rp [ppm] & $\mathcal{U}$(0,4000) & $1638^{+40}_{-28}$ \\
LC jitter LCOHAL_zs [ppm] & $\mathcal{U}$(0,4000) & $1164^{+60}_{-38}$ \\
LC jitter MEarth [ppm] & $\mathcal{U}$(0,4000) & $881^{+28}_{-41}$ \\

\hline
\textit{GP parameters} & &  \\
\hline
$\eta_{\sigma,LC}$ [ppm] & $\mathcal{LU}$(0.01,10000) & $2620^{+600}_{-410}$ \\
$\eta_{{\rm Q}_0}$ & $\mathcal{U}$(-8.0,10) & $-0.82^{+0.17}_{-0.17}$ \\
$\eta_{\rho}$ [days] & $\mathcal{G}$(24.0,1) & $23.97^{+1.04}_{-0.92}$ \\
$\eta_{\delta{\rm Q}}$ & $\mathcal{U}$(0.13,7.38) & $3.5^{+3.1}_{-2.8}$ \\
$\eta_{f}$ & $\mathcal{U}$(0,1) & $0.71^{+0.26}_{-0.20}$ \\

\hline
\hline

\multicolumn{3}{l}{Notes: $\mathcal{G}(\mu,\sigma)$: Normal distribution with mean $\mu$ and width $\sigma$. $\mathcal{U}(a,b)$: Uniform distribution between $a$ and $b$.}\\
\multicolumn{3}{l}{$\mathcal{LU}(a,b)$: Log-uniform distribution between $a$ and $b$. $\mathcal{T}(\mu,\sigma,a,b)$: Truncated normal distribution with mean $\mu$ and width $\sigma$,}\\
\multicolumn{3}{l}{between $a$ and $b$. }\\


\end{tabular}
\end{table*}


\begin{table*}
\setlength{\extrarowheight}{3pt}
\centering
\caption{Inferred and derived parameters from the radial velocity only analysis (Sect.~\ref{sec:RVanalysis}) for the preferred model (labelled as "2p1cF") with the inner planet in a circular orbit and the external component in an eccentric orbit.}
\label{tab:RValone}
\begin{tabular}{lll}
\hline\hline
Parameter & Priors & Posteriors \\
\hline
\textit{Planet b} & &  \\
\hline
Orbital period, $P_b$ [days] & $\mathcal{G}$(1.8237371,0.000012) & $1.823737^{+0.000013}_{-0.000012}$ \\
Time of mid-transit, $T_{\rm 0,b}-2400000$ [days] & $\mathcal{G}$(59248.377773,0.003) & $59248.3784^{+0.0031}_{-0.0031}$ \\
RV semi-amplitude, $K_{\rm b}$ [m/s] & $\mathcal{U}$(0.0,100.0) & $5.75^{+0.67}_{-0.62}$ \\
Planet mass, $m_{b}\sin{i_b}$ [\Mearth{}] & (derived) & $8.89^{+1.0}_{-0.95}$ \\
Orbit semi-major axis, $a_{b}$ [AU] & (derived) & $0.02635^{+0.00017}_{-0.00017}$ \\
Relative orbital separation, $a_{b}/R_{\star}$ & (derived) & $8.44^{+0.20}_{-0.19}$ \\
Incident Flux, $F_{\rm inc,b}$ [$F_{{\rm inc},\oplus}$] & (derived) & $225^{+20}_{-19}$ \\

\hline
\textit{Planet c} & &  \\
\hline
Orbital period, $P_c$ [days] & $\mathcal{LU}$(100.0,5000.0) & $1730^{+240}_{-270}$ \\
Time of mid-transit, $T_{\rm 0,c}-2400000$ [days] & $\mathcal{U}$(57200.0,62900.0) & $(60670^{+220}_{-250}$ \\
RV semi-amplitude, $K_{\rm c}$ [m/s] & $\mathcal{U}$(0.0,1500.0) & $319^{+23}_{-20}$ \\
Orbital eccentricity, $e_{\rm c}$ & $\mathcal{U}$(0.0,1.0) & $0.622^{+0.035}_{-0.040}$ \\
Arg. periastron, $\omega_{\rm c}$ [deg.] & $\mathcal{U}$(0.0,360.0) & $208.4^{+6.6}_{-6.4}$ \\
Planet mass, $m_{c}\sin{i_c}$ [\Mjup{}] & (derived) & $11.8^{+1.6}_{-1.1}$ \\
Orbit semi-major axis, $a_{c}$ [AU] & (derived) & $2.54^{+0.38}_{-0.36}$ \\
Relative orbital separation, $a_{c}/R_{\star}$ & (derived) & $810 \pm 120$ \\
Incident Flux, $F_{\rm inc,c}$ [$F_{{\rm inc},\oplus}$] & (derived) & $0.0243^{+0.0089}_{-0.0062}$ \\

\hline
\textit{Instrument-dependent parameters} & &  \\
\hline

$\delta_{\rm HARPS}$ [km/s] & $\mathcal{U}$(-10,10) & $-5.479^{+0.021}_{-0.025}$ \\
$\delta_{\rm CORALIE}$ [km/s] & $\mathcal{U}$(-10,10) & $-5.541^{+0.018}_{-0.023}$ \\
$\delta_{\rm PFS}$ [km/s] & $\mathcal{U}$(-10,10) & $0.178^{+0.021}_{-0.025}$ \\
$\sigma_{\rm HARPS}$ [m/s] & $\mathcal{LU}$(0.1,2) & $1.72^{+0.21}_{-0.43}$ \\
$\sigma_{\rm CORALIE}$ [m/s] & $\mathcal{LU}$(0.1,30) & $1.20^{+7.7}_{-0.99}$ \\
$\sigma_{\rm PFS}$ [m/s] & $\mathcal{LU}$(0.1,5) & $3.74^{+0.88}_{-1.1}$ \\
$\delta_{\rm FWHM,HARPS}$ [km/s] & $\mathcal{U}$(-0.5,0.5) & $0.0575^{+0.0017}_{-0.0016}$ \\
$\delta_{\rm FWHM,CORALIE}$ [km/s] & $\mathcal{U}$(-0.5,0.5) & $-0.077^{+0.013}_{-0.014}$ \\
$\sigma_{\rm FWHM,HARPS}$ [m/s] & $\mathcal{LU}$(0.1,300) & $6.4^{+1.7}_{-1.3}$ \\
$\sigma_{\rm FWHM,CORALIE}$ [m/s] & $\mathcal{LU}$(0.1,300) & $61.7^{+11}_{-7.4}$ \\

\hline
\textit{GP parameters} & &  \\
\hline
$\eta_{\rm 1,FWHM}$ [m/s] & $\mathcal{LU}$(0.01,400) & $6.2^{+2.4}_{-4.4}$ \\
$\eta_1$ [m/s] & $\mathcal{LU}$(0.01,50) & $9.3^{+1.8}_{-1.4}$ \\
$\eta_2$ [days] & $\mathcal{U}$(-8.0,10) & $-2.20^{+0.35}_{-0.37}$ \\
$\eta_3$ [days] & $\mathcal{G}$(24.0,1) & $24.06^{+0.95}_{-1.0}$ \\
$\eta_4$ & $\mathcal{U}$(0.13,7.38) & $3.9^{+2.4}_{-2.6}$ \\
\hline
\hline

\multicolumn{3}{l}{Notes:}\\
\multicolumn{3}{l}{$\bullet$ $\mathcal{G}(\mu,\sigma)$: Normal distribution with mean $\mu$ and width $\sigma$}\\
\multicolumn{3}{l}{$\bullet$ $\mathcal{U}(a,b)$: Uniform distribution between $a$ and $b$}\\
\multicolumn{3}{l}{$\bullet$ $\mathcal{LU}(a,b)$: Log-uniform distribution between $a$ and $b$}\\

\end{tabular}
\end{table*}


\onecolumn

\setlength{\extrarowheight}{3pt}
\begin{longtable}{lcc}
\caption{\label{tab:joint} Inferred and derived parameters from the joint radial velocity and light curves analysis (Sect.~\ref{sec:joint}) for the preferred model.}\\
\hline
Parameter & Priors & Posteriors \\
\hline
\endfirsthead
\multicolumn{3}{l}{{\bfseries \tablename\ \thetable{} -- continued from previous page}} \\
\hline
Parameter & Priors & Posteriors \\
\hline
\endhead
\multicolumn{3}{l}{{Continued on next page}} \\ 
\hline
\endfoot
\hline
\multicolumn{3}{l}{Notes:}\\
\multicolumn{3}{l}{$\bullet$ $\mathcal{G}(\mu,\sigma)$: Normal distribution with mean $\mu$ and width $\sigma$}\\
\multicolumn{3}{l}{$\bullet$ $\mathcal{U}(a,b)$: Uniform distribution between $a$ and $b$}\\
\multicolumn{3}{l}{$\bullet$ $\mathcal{LU}(a,b)$: Log-uniform distribution between $a$ and $b$}\\
\multicolumn{3}{l}{$\bullet$ $\mathcal{T}(\mu,\sigma,a,b)$: Truncated normal distribution with mean $\mu$ and width $\sigma$, between $a$ and $b$}\\
\endlastfoot
\multicolumn{3}{l}{\it Planet b}  \\
\hline
Orbital period, $P_b$ [days] & $\mathcal{G}$(1.8237371,0.0001) & $1.8237305^{+0.0000020}_{-0.0000021}$ \\
Time of mid-transit, $T_{\rm 0,b}-2400000$ [days] & $\mathcal{G}$(59248.377773,0.1) & $59248.37709^{+0.00036}_{-0.00039}$ \\
Planet mass, $M_b$ [\Mearth{}] & $\mathcal{U}$(0,5000) & $9.1^{+1.0}_{-1.0}$ \\
Orbital inclination, $i_{\rm b}$ [deg.] & $\mathcal{U}$(40,90) & $86.75^{+0.38}_{-0.41}$ \\
Planet radius, $R_b$ [\Rearth{}] & $\mathcal{U}$(0,20) & $2.765^{+0.088}_{-0.097}$ \\
Planet density, $\rho_{b}$ [$g\cdot cm^{-3}$] & (derived) & $2.34^{+0.39}_{-0.34}$ \\
Transit depth, $\Delta_{b}$ [ppt] & (derived) & $1.435^{+0.043}_{-0.043}$ \\
Orbit semi-major axis, $a_{b}$ [AU] & (derived) & $0.02636^{+0.00017}_{-0.00017}$ \\
Relative orbital separation, $a_{b}/R_{\star}$ & (derived) & $8.47^{+0.21}_{-0.21}$ \\
Transit duration, $T_{\rm 14,b}$ [hours] & (derived) & $1.519^{+0.018}_{-0.018}$ \\
Planet surface gravity, $g_{\rm b}$ [$m \cdot s^{-2}$] & (derived) & $11.6^{+1.6}_{-1.5}$ \\
Impact parameter, $b_{b}/R_{\star}$ & (derived) & $0.480^{+0.044}_{-0.051}$ \\
Incident Flux, $F_{\rm inc,b}$ [$F_{{\rm inc},\oplus}$] & (derived) & $188^{+26}_{-24}$ \\
Equilibrium temperature, $T_{\rm eq,b}$ [K] & (derived) & $941^{+31}_{-31}$ \\

\hline
\multicolumn{3}{l}{\it Planet c}  \\
\hline
Orbital period, $P_c$ [days] & $\mathcal{LU}$(100,5000) & $1700^{+290}_{-280}$ \\
Time of mid-transit, $T_{\rm 0,c}-2400000$ [days] & $\mathcal{U}$(57200,62900) & $60640^{+260}_{-260}$ \\
Minimum mass, $m_c\sin{i_c}$ [\Mjup{}] & $\mathcal{U}$(0,30) & $11.3^{+1.1}_{-0.9}$ \\
Orbital eccentricity, $e_{\rm c}$ & $\mathcal{U}$(0,1) & $0.628^{+0.043}_{-0.036}$ \\
Arg. periastron, $\omega_{\rm c}$ [deg.] & $\mathcal{U}$(0,360) & $208.5^{+7.8}_{-7.3}$ \\
Orbit semi-major axis, $a_{c}$ [AU] & (derived) & $2.52^{+0.27}_{-0.29}$ \\
Relative orbital separation, $a_{c}/R_{\star}$ & (derived) & $806^{+91}_{-93}$ \\
Incident Flux, $F_{\rm inc,c}$ [$F_{{\rm inc},\oplus}$] & (derived) & $0.0208^{+0.0065}_{-0.0046}$ \\
Equilibrium temperature, $T_{\rm eq,c}$ [K] & (derived) & $96.4^{+6.8}_{-5.8}$ \\

\hline
\multicolumn{3}{l}{\it Stellar parameters}  \\
\hline
Stellar radius, $R_{\star}$ [$R_{\odot}$] & $\mathcal{T}$(0.671,0.015,0.1,1.2) & $0.669^{+0.015}_{-0.016}$ \\
Stellar mass, $M_{\star}$ [$M_{\odot}$] & $\mathcal{T}$(0.734,0.014,0.1,1.2) & $0.735^{+0.014}_{-0.015}$ \\
Stellar luminosity, $L_{\star}$ [$L_{\odot}$] & (derived) & $0.368^{+0.018}_{-0.018}$ \\

Limb-darkening $u_1$, LCOHAL_gp & $\mathcal{T}$(0.592,0.05,0,1) & $0.602^{+0.050}_{-0.048}$ \\
Limb-darkening $u_2$, LCOHAL_gp & $\mathcal{T}$(0.141,0.05,0,1) & $0.142^{+0.049}_{-0.048}$ \\
Limb-darkening $u_1$, LCOHAL_ip & $\mathcal{T}$(0.539,0.05,0,1) & $0.534^{+0.050}_{-0.049}$ \\
Limb-darkening $u_2$, LCOHAL_ip & $\mathcal{T}$(0.155,0.05,0,1) & $0.150^{+0.051}_{-0.050}$ \\
Limb-darkening $u_1$, LCOHAL_rp & $\mathcal{T}$(0.7,0.05,0,1) & $0.728^{+0.051}_{-0.049}$ \\
Limb-darkening $u_2$, LCOHAL_rp & $\mathcal{T}$(0.078,0.05,0,1) & $0.102^{+0.047}_{-0.048}$ \\
Limb-darkening $u_1$, LCOHAL_zs & $\mathcal{T}$(0.416,0.05,0,1) & $0.411^{+0.051}_{-0.050}$ \\
Limb-darkening $u_2$, LCOHAL_zs & $\mathcal{T}$(0.195,0.05,0,1) & $0.189^{+0.048}_{-0.051}$ \\
Limb-darkening $u_1$, MEarth & $\mathcal{T}$(0.416,0.1,0,1) & $0.465^{+0.094}_{-0.094}$ \\
Limb-darkening $u_2$, MEarth & $\mathcal{T}$(0.195,0.1,0,1) & $0.217^{+0.095}_{-0.095}$ \\

\hline
\multicolumn{3}{l}{\it Instrument-dependent parameters}  \\
\hline
LC level TESS07 & $\mathcal{U}$(-20,20) & $-0.71^{+0.79}_{-0.74}$ \\
LC level TESS34 & $\mathcal{U}$(-20,20) & $-0.12^{+0.77}_{-0.77}$ \\
LC level LCOHAL_gp & $\mathcal{U}$(-200,200) & $5.7^{+0.17}_{-0.18}$ \\
LC level LCOHAL_ip & $\mathcal{U}$(-200,200) & $-0.47^{+0.21}_{-0.21}$ \\
LC level LCOHAL_rp & $\mathcal{U}$(-200,200) & $0.64^{+0.19}_{-0.20}$ \\
LC level LCOHAL_zs & $\mathcal{U}$(-200,200) & $-1.25^{+0.20}_{-0.20}$ \\
LC level MEarth & $\mathcal{U}$(-200,200) & $5.42^{+0.18}_{-0.19}$ \\
LC jitter TESS07 [ppm] & $\mathcal{U}$(0,4000) & $98^{+42}_{-29}$ \\
LC jitter TESS34 [ppm] & $\mathcal{U}$(0,4000) & $81^{+51}_{-45}$ \\
LC jitter LCOHAL_gp [ppm] & $\mathcal{U}$(0,4000) & $871^{+40}_{-40}$ \\
LC jitter LCOHAL_ip [ppm] & $\mathcal{U}$(0,4000) & $1575^{+42}_{-42}$ \\
LC jitter LCOHAL_rp [ppm] & $\mathcal{U}$(0,4000) & $1643^{+38}_{-37}$ \\
LC jitter LCOHAL_zs [ppm] & $\mathcal{U}$(0,4000) & $1177^{+52}_{-52}$ \\
LC jitter MEarth [ppm] & $\mathcal{U}$(0,4000) & $877^{+38}_{-39}$ \\
$\delta_{\rm HARPS}$ [km/s] & $\mathcal{U}$(-10,10) & $-5.478^{+0.025}_{-0.025}$ \\
$\delta_{\rm CORALIE}$ [km/s] & $\mathcal{U}$(-10,10) & $-5.537^{+0.021}_{-0.020}$ \\
$\delta_{\rm PFS}$ [km/s] & $\mathcal{U}$(-10,10) & $0.178^{+0.025}_{-0.025}$ \\
$\sigma_{\rm HARPS}$ [m/s] & $\mathcal{LU}$(0.1,2) & $1.51^{+0.74}_{-0.37}$ \\
$\sigma_{\rm CORALIE}$ [m/s] & $\mathcal{LU}$(0.1,30) & $9.9^{+6.4}_{-7.8}$ \\
$\sigma_{\rm PFS}$ [m/s] & $\mathcal{LU}$(0.1,5) & $4.0^{+1.2}_{-0.7}$ \\

\hline
\multicolumn{3}{l}{\it RV and LC Gaussian process hyper-parameters}  \\
\hline
$\eta_{\rm 1,LC}$ [ppm] & $\mathcal{LU}$(10000) & $2440^{+780}_{-220}$ \\
$\eta_{\rm 1,FWHM}$ [m/s] & $\mathcal{LU}$(0.01,400) & $23.2^{+4.2}_{-5.2}$ \\
$\eta_1$ [m/s] & $\mathcal{LU}$(0.01,50) & $11.6^{+1.6}_{-2.0}$ \\
$\eta_2$ [days] & $\mathcal{U}$(-8.0,10) & $-0.71^{+0.19}_{-0.22}$ \\
$\eta_3$ [days] & $\mathcal{G}$(24,1) & $24.6^{+1.35}_{-0.87}$ \\
$\eta_4$ & $\mathcal{U}$(0.13,7.38) & $1.59^{+0.87}_{-1.22}$ \\

\end{longtable}

\end{document}